\documentclass[preprint,11pt]{elsarticle}

\usepackage{graphics}
\usepackage{graphicx}
\usepackage{epsfig}

\usepackage{amssymb}
\usepackage{epstopdf}
\usepackage{amsmath}
\usepackage{bm}
\usepackage[T1]{fontenc}
\usepackage{textcomp}
\usepackage{color}

\journal{Advances in Colloid and Interface Science}

\begin{document}

\begin{frontmatter}



\title{Physical aspects of heterogeneities in multi-component lipid membranes}


\author{Shigeyuki Komura}
\ead{komura@tmu.ac.jp}
\address{Department of Chemistry,
Graduate School of Science and Engineering, \\
Tokyo Metropolitan University, Tokyo 192-0397, Japan}

\author{David Andelman\corref{cor1}}
\ead{andelman@post.tau.ac.il}
\address{Raymond and Beverly Sackler School of Physics and Astronomy, \\
Tel Aviv University, Ramat Aviv, Tel Aviv 69978, Israel}
\cortext[cor1]{Corresponding author}

\date{Ver 7 - 15/11/2013}

\begin{abstract}
Ever since the raft model for biomembranes has been proposed, the traditional view
of biomembranes based on the fluid-mosaic model has been altered.
In the raft model, dynamical heterogeneities in multi-component lipid
bilayers play an essential role.
Focusing on the lateral phase separation of biomembranes and vesicles, we review
some of the most relevant research conducted over the last decade.
We mainly refer to those experimental works that are based on physical
chemistry approach, and to theoretical explanations given in terms of
soft matter physics.
In the first part, we describe the phase behavior and the conformation of
multi-component lipid bilayers.
After formulating the hydrodynamics of fluid membranes in presence of
the surrounding solvent, we discuss the domain growth-law and decay rate of
concentration fluctuations.
Finally, we review several attempts to describe membrane rafts as two-dimensional microemulsion.
\end{abstract}

%\begin{keyword}
%keyword1 \sep keyword2


%\end{keyword}

\end{frontmatter}

\tableofcontents

% \linenumbers


%%%%%%%%%%%%%%%%%%%%%%%%%%%%%%%%%%%%%%%%%%%%%%%%%%%%%%%%%%%%%%%%%%%%%%%%%
\section{Introduction}
\label{sec:introduction}
%%%%%%%%%%%%%%%%%%%%%%%%%%%%%%%%%%%%%%%%%%%%%%%%%%%%%%%%%%%%%%%%%%%%%%%%%

Biomembranes, which delimit the boundaries of biological cells, as well as the perimeter of
intra-cellular organelles,
play a vital role in maintaining and regulating cellular functions~\cite{alberts}.
For example, mitochondria and chloroplast are examples of specific 
cellular organelles that use  concentration gradient
of ions across their membrane to supply energy that is indispensable
to biological activities.

The main building blocks of biomembranes are a variety of phospholipids, glycosphingolipids and cholesterol.
Phospholipids and glycolipids are amphiphilic molecules consisting of two moieties:
a hydrophilic head group and, typically, two hydrophobic hydrocarbon chains.
When lipid molecules are dissolved in water, they spontaneously form a bilayer
membrane, where the hydrocarbon tails of the two leaflets face each other and point away from the water phase,
in order to avoid direct contact between the hydrophobic hydrocarbon chains and
water.
In addition,  various types of proteins are embedded within
the biomembrane, and among other roles, they mediate the transport of substances between the
inside and  outside of the cell.

At physiological temperatures, biomembranes are fluid. This guarantees that both lipids and proteins
can freely diffuse laterally within the membrane plane.
A dynamical picture of biomembranes was provided by Singer and
Nicolson in 1972, and their model was named the {\it "fluid mosaic
model"}~\cite{SN72}.
Although this model became a widely accepted standard model of
biomembranes, the so-called {\it "raft model"} proposed by Simons and Ikonen~\cite{simons-97} in 1997
initiated a substantial debate in the community regarding the nature of structural heterogeneities
of biomembranes and their implied function.

According to the fluid mosaic model, various lipids and proteins are considered to be
uniformly distributed within the membrane plane, while the raft model asserts
the existence of nanoscale domains consisting of cholesterol and specific phospholipids such as
sphingolipid.
These lipid domains together with membrane proteins are expected to act as relay
stations for signaling transfer factor, and to be involved in controlling 
quite a number of biological processes.

Based on  the accumulated experimental results, a definition of
lipid rafts has been suggested in an international conference held
in 2006~\cite{pike06}:
\begin{quotation}
Membrane rafts are small (10--200 nm), heterogeneous, highly dynamic, sterol
and sphingolipid-enriched domains that compartmentalize cellular processes.
Small rafts can sometimes be stabilized to form larger platforms through
protein-protein and protein-lipid interactions.
\end{quotation}
However, even today, there are still a large number of unresolved questions
with regard to the origin and actual nature of membrane rafts {\it in vivo} 
and {\it in vitro}~\cite{leslie11}.
One of the reasons why the scientific community has been unable to settle the issue
concerning the raft hypothesis is due to a lack of direct visualization of
rafts in bio-membranes.
Nevertheless, the raft hypothesis has largely stimulated the interest
of physicists and physical chemists, leading to many studies on heterogeneities and
phase separation of model lipid membranes.

In this article, we review the physical phenomenon that is induced by
phase separation in multi-component lipid membranes, and describe related
experimental and theoretical studies.
However, it should be noted with care that exactly how the observed structures in model
bilayers are related to lipid rafts in biomembranes is not completely understood at present.
Nevertheless, the real picture of lipid rafts may become more
evident as we explore physical mechanisms and concepts such as diffusion, phase
separation and critical phenomena, because those mechanisms are well-defined and can be
quantitatively measured and investigated in model
systems.

The outline of this article is as follow. In the next section, we explain some known facts concerning
lateral phase separation in multi-component membranes and review several physical
models that describe it.
In Sec.~\ref{sec:vesicle}, we discuss the coupling between the lateral phase
separation with membrane deformation and vesicle shape.
Section \ref{sec:hydro} deals with hydrodynamics of fluid membranes
embedded in a surrounding solvent.
Based on hydrodynamic arguments, we review the dynamics of lateral phase
separation in Sec.~\ref{sec:separation}, particularly focusing on
domain growth-law below the transition temperature.
In Sec.~\ref{sec:fluctuation}, we summarize several results
on the dynamics of concentration fluctuations above the transition temperature, and
Sec.~\ref{sec:microemulsion} is concerned with  several recent attempts to view 
membrane rafts as two-dimensional microemulsion.
Finally, an outlook is provided in the last section.

%%%%%%%%%%%%%%%%%%%%%%%%%%%%%%%%%%%%%%%%%%%%%%%%%%%%%%%%%%%%%%%%%%%%%%%%%
\section{Lateral phase separation in multi-component membranes}
\label{sec:phase}
%%%%%%%%%%%%%%%%%%%%%%%%%%%%%%%%%%%%%%%%%%%%%%%%%%%%%%%%%%%%%%%%%%%%%%%%%
\subsection{Structural phase-transition of lipid bilayers}
%%%%%%%%%%%%%%%%%%%%%%%%%%%%%%%%%%%%%%%%%%%%%%%%%%%%%%%%%%%%%%%%%%%%%%%%%

Lipids, which are the building blocks of biomembranes, include phospholipids, glycosphingolipids,
and cholesterol.
Typical examples of phospholipids are phosphatidylcholine (PC),
phosphatidylserine (PS), and sphingomyelin (SM).
Below, we shall divide lipid molecules into two categories: {\it "saturated lipids"}
which do not contain any unsaturated bond in their hydrocarbon chains,
and {\it "unsaturated lipids"} which have at least one unsaturated bond.
Although cholesterol is also a lipid, hereafter
we shall refer only to phospholipids or glycosphingolipids as {\it "lipids"} and
cholesterol will retain its name.

It is known that by changing temperature, single-component lipid bilayers
will undergo a structural phase transition that
reflects a change in the orientational order of the lipid hydrocarbon
chains~\cite{lipowsky-sackmann}.
At high temperatures, lipid bilayers are in a {\it "liquid-crystalline phase"},
whereas at low temperatures they are in a {\it "gel phase"}, in which
the lipid molecules hardly diffuse.
Prior to the lipid raft hypothesis, the phase behavior of bilayers
consisting of two types of lipids characterized by different transition
temperatures has been investigated using various experimental
methods~\cite{lipowsky-sackmann}.
Since saturated lipids typically have higher phase-transition temperature
than unsaturated ones, a region of two-phase coexistence appears
between the liquid-crystalline and gel phases over a certain
temperature range.

Cholesterol, on the other hand, is known to exhibit a dual effect in lipid
bilayers~\cite{IKMWZ}.
In the liquid-crystalline phase cholesterol promotes the
packing of hydrocarbon chains, while it disrupts the chain ordering in the
gel phase.
In binary mixtures of lipid and cholesterol, a mixed phase
called the {\it "liquid-ordered phase"} (L$_{\rm o}$-phase) appears when
cholesterol concentration is high enough.
In this phase, even though the lipid hydrocarbon chains are relatively ordered,
the membrane maintains its fluidity.
For intermediate cholesterol concentrations, membranes undergo a liquid-liquid
phase separation between the L$_{\rm o}$-phase and a 
lipid-rich phase called the {\it "liquid-disordered phase"} (L$_{\rm d}$-phase). The latter
phase is identical to the previously introduced {\it "liquid-crystalline phase"},
and the lipid hydrocarbon chains in this phase are less ordered as compared
to the L$_{\rm o}$-phase.
Since the minority phase forms domains as a result of phase separation, these domains have
been studied in hope to shed light on membrane rafts 
with whom they probably share some similarities.

%%%%%%%%%%%%%%%%%%%%%%%%%%%%%%%%%%%%%%%%%%%%%%%%%%%%%%%%%%%%%%%%%%%%%%%%%
\subsection{Three-component lipid mixtures}
%%%%%%%%%%%%%%%%%%%%%%%%%%%%%%%%%%%%%%%%%%%%%%%%%%%%%%%%%%%%%%%%%%%%%%%%%

\begin{figure}[b]
\begin{center}
\resizebox{1.0\columnwidth}{!}{
\includegraphics{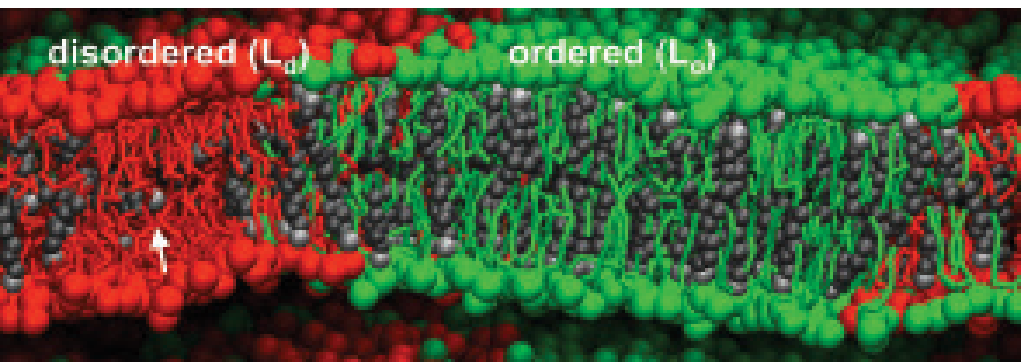}
}
\caption{
Coarse-grained molecular dynamics simulation of a three-component lipid bilayer
membrane showing lateral phase separation.
Green molecules are saturated lipids, red molecules are unsaturated lipids,
and black molecules are cholesterol.
The shown membrane is in a two-phase coexisting region between the L$_{\rm o}$-phase (right) and
the L$_{\rm d}$-phase (left).
The white arrow points to a cholesterol oriented in between the monolayer
leaflets.
Adapted from Ref.~\cite{RM}. 
}
\label{fig:marrink}
\end{center}
\end{figure}

Dietrich \textit{et al.}~\cite{Dietrich} were the first to visualize liquid domains in
three-component lipid bilayers.
Using fluorescence microscope, they observed phase-separated patterns in giant
vesicles composed of unsaturated lipid, saturated lipid, and cholesterol.
They demonstrated that the liquid-liquid phase separation indeed occurs because
the L$_{\rm o}$-phase rich in saturated lipid and cholesterol form circular
two-dimensional (2d) liquid domains.
Following this work, several other studies have been
conducted~\cite{VK02,veatch-03,veatch-05} elucidating  the phase behavior 
of various combinations of three-component
mixtures (unsaturated lipid/saturated lipid/ cholesterol) as a function
of temperature and composition.
We note that a typical and well-studied three-component
mixture is that of
DOPC\footnote{DOPC: dioleoyl-phosphatidylcholine
(unsaturated lipid)}/DPPC\footnote{DPPC:
dipalmitoyl-phosphatidylcholine
(saturated lipid)}/ cholesterol.
In Fig.~\ref{fig:marrink}, we show a visualization obtained by
coarse-grained molecular dynamics simulation~\cite{RM} of a three-component
lipid bilayer membrane exhibiting a lateral phase separation between 
L$_{\rm o}$ and L$_{\rm d}$ phases. 

For our later discussion, we show in Fig.~\ref{fig:ternaryphase}
the experimentally obtained~\cite{VGK06} ternary phase diagram of 
diPhyPC\footnote{diPhyPC: diphytanoyl-phosphatidylcholine. 
Although this is a saturated lipid, it has a very low structural 
phase-transition temperature because of the branched structure of the 
hydrocarbon chains.
Hence, it plays a similar role to unsaturated lipids.}/DPPC/ cholesterol 
for two different temperatures: 
43\,$^\circ$C (top) and 16\,$^\circ$C (bottom).
Fluorescence microscope images of vesicles are also shown for different
compositions.
In these images, the white regions are rich in diPhyPC, while the dark
ones are rich in DPPC and cholesterol.
The open circles in the ternary phase diagram correspond to 
one-phase region (homogeneous phase),
the filled circles to the two-phase coexisting region between
the L$_{\rm o}$-phase and the L$_{\rm d}$-phase, and
the gray squares indicate the gel phase.
Interestingly, the two-phase coexistence region forms a closed loop at the
higher temperature (43\,$^{\circ}$C).
At the lower temperature (16\,$^{\circ}$C), two-phase coexistence region
can be seen between the L$_{\rm d}$-phase and the gel phase, 
below the triangular coexistence region of the three phases.

%%%%%%%%%%%%%%%%%%%%%%%%%%%%%%%%%%%%%%%%%%%%%%%%%%%%%%%%%%%%%%%%%%%%%%%%%
\subsection{Theoretical models for lipid mixtures}
%%%%%%%%%%%%%%%%%%%%%%%%%%%%%%%%%%%%%%%%%%%%%%%%%%%%%%%%%%%%%%%%%%%%%%%%%

There have been several theoretical attempts to predict and reproduce the phase behavior
of multi-component lipid membranes.
The membrane structural phase transition is generally first order, and is
analogous to the nematic-isotropic transition in liquid crystals.
Hence, the structural phase-transition of each type of lipid can be described
by the Landau--de Gennes free-energy~\cite{deGennes}. A model for
biomembranes consisting of two types of lipids having different
structural phase-transition temperature was proposed by Komura
\textit{et al.}~\cite{KSOA,KSO}.
The phase-transition temperature of the binary lipid membrane was assumed to be
a linear interpolation between the transition temperatures
of the two pure components, and
the model has been successful in explaining general phase behavior of
various combinations of binary lipid mixtures.

As for lipid and cholesterol binary mixtures, by focusing on the dual effects
of cholesterol as mentioned above, Ipsen \textit{et al.}~\cite{IKMWZ} proposed a microscopic
model, while Komura \textit{et al.}~\cite{KSOA} developed a phenomenological
model.
It should be noted, however, that role of cholesterol is not yet fully understood.

Next, we review models addressing the phase behavior of three-component
membranes in which cholesterol is added to a binary lipid mixture.
For example, by using a self-consistent molecular model for cholesterol and lipids,
Elliot \textit{et al.}~\cite{ESS} calculated the phase diagram of ternary mixtures.
Several phenomenological models have been also
proposed~\cite{PS08,WMT},
among them, we briefly explain the model proposed by Putzel \textit{et al.}~\cite{PS08}.
In that work the area fraction (related to the relative concentration)
of unsaturated lipid, saturated lipid, and
cholesterol is denoted by $\phi_{\rm u}$, $\phi_{\rm s}$, and $\phi_{\rm c}$,
respectively, and satisfies the incompressibility condition
$\phi_{\rm u}+\phi_{\rm s}+\phi_{\rm c}=1$.
In addition, a parameter $\delta$ is associated with the saturated lipids 
and characterizes the degree
of orientational order of their hydrocarbon chains;
the larger the value of $\delta$, the more orientational order of the hydrocarbon
chains\footnote{Since $\delta$ is not a rigorous order parameter, its absolute
value does not have any physical meaning.}.
The free-energy density $f_{\rm \ell}$ of the liquid phase was given by~\cite{PS08}
\begin{align}
f_{\rm \ell} & = J_{\rm ss} \phi_{\rm s}^2 (\delta-1)^2
+J_{\rm us} \phi_{\rm u} \phi_{\rm s} \delta
-J_{\rm cs} \phi_{\rm c} \phi_{\rm s} \delta(1 - \delta),\nonumber\\
&+ k_{\rm B}T (\phi_{\rm u} \ln \phi_{\rm u}
+ \phi_{\rm s} \ln \phi_{\rm s} + \phi_{\rm c} \ln \phi_{\rm c} )
\label{putzel}
\end{align}
where $k_{\rm B}$ is the Boltzmann constant, $T$ the temperature, and
$J_{\rm ss}$, $J_{\rm us}$, $J_{\rm cs}$ are all positive interaction parameters.
The  $J_{\rm ss}$ term represents the interaction between
two saturated lipids and depends on the orientational order $\delta$.
In a similar way, the $J_{\rm us}$ and $J_{\rm cs}$ terms
correspond to unsaturated-saturated lipids and cholesterol-saturated lipid
interactions, respectively.
The last three terms account for the ideal entropy of mixing.

\begin{figure}[t]
\begin{center}
\resizebox{0.9\columnwidth}{!}{
\includegraphics{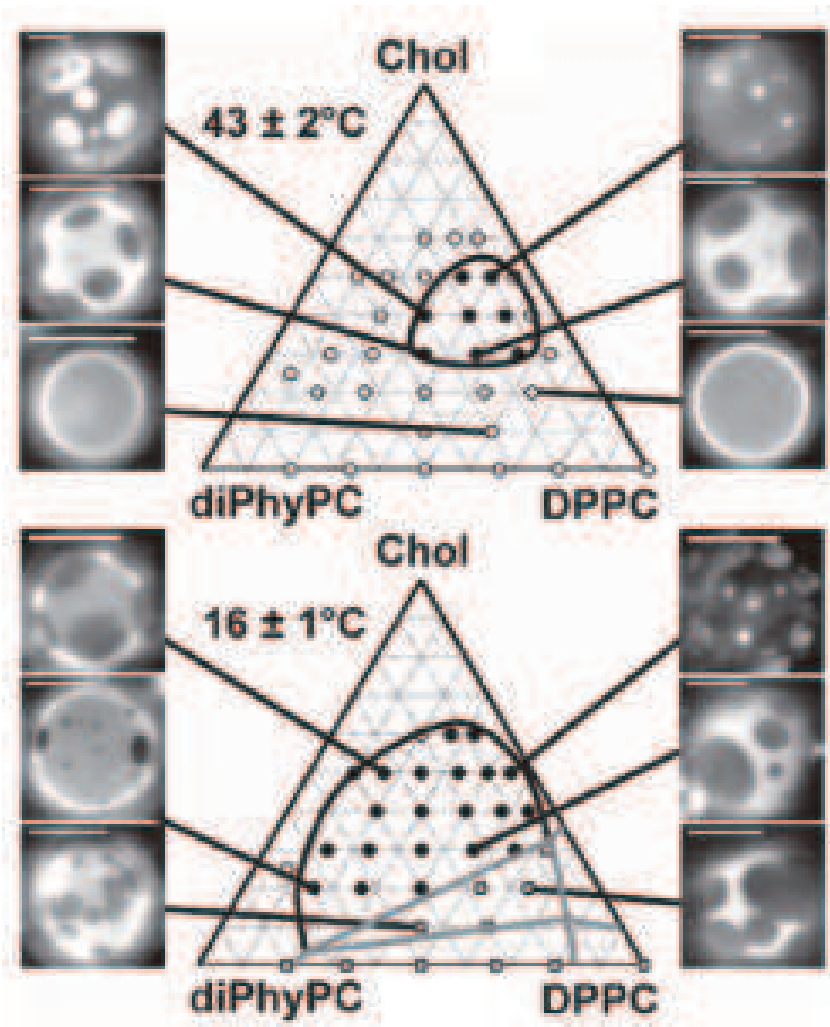}
}
\caption{
Fluorescence microscope images and ternary phase diagrams of giant
vesicles composed of diPhyPC/DPPC/ cholesterol at 43\,$^{\circ}$\,C (top)
and 16\,$^{\circ}$\,C (bottom).
Dark circular domains in the side images are rich in DPPC and cholesterol.
Open circles reside in the  one-phase region of the phase diagrams,
filled circles correspond to the two-phase coexisting region between the
L$_{\rm o}$-phase and the L$_{\rm d}$-phase (liquid-liquid phase
separation), and the gray squares indicate the gel phase.
The scale bar corresponds to 20 $\mu$m.
Adapted from Ref.~\cite{VGK06}.
}
\label{fig:ternaryphase}
\end{center}
\end{figure}

\begin{figure}[t]
\begin{center}
\resizebox{0.65\columnwidth}{!}{
\includegraphics{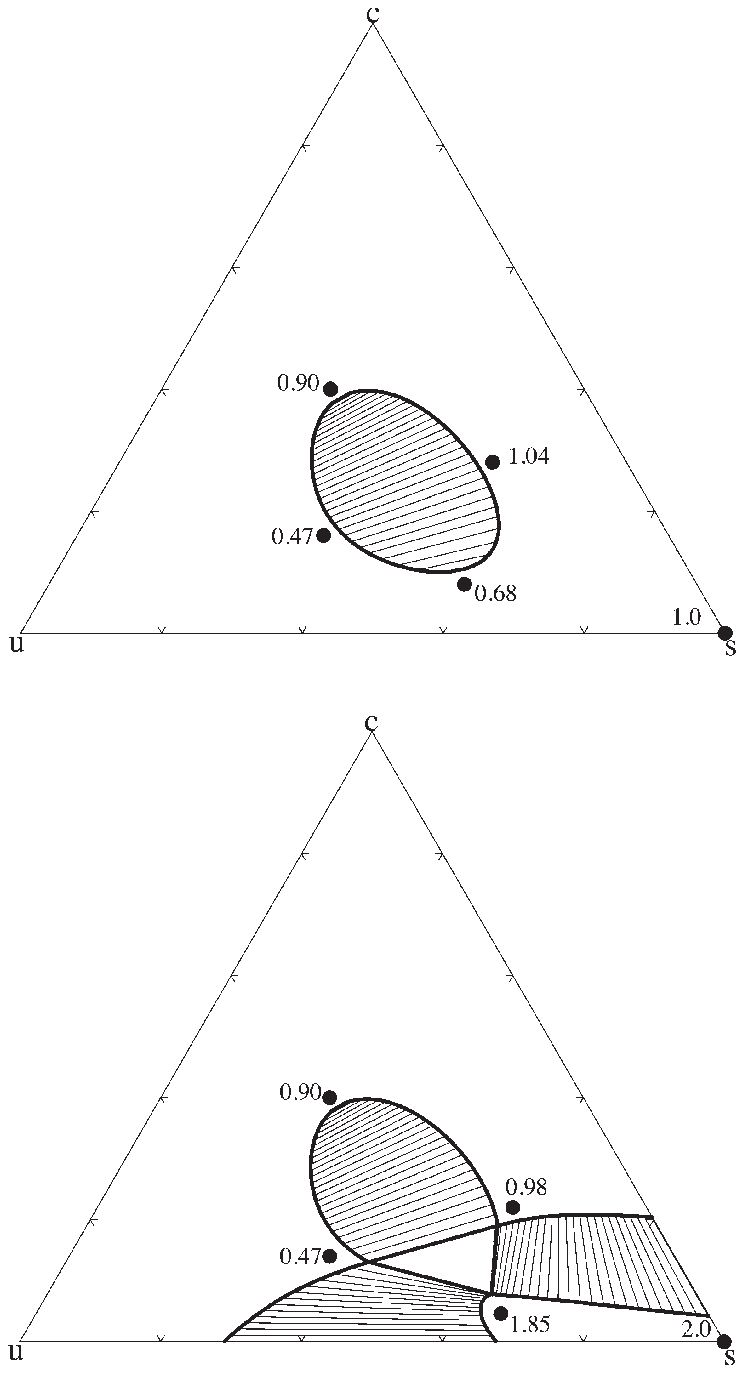}
}
\caption{
Ternary phase diagram of saturated lipid, unsaturated lipid and cholesterol, 
as obtained from the free energy in Eq.~(\ref{putzel}).
The triangular corners "u", "s", and "c" represent unsaturated lipid,
saturated lipid, and cholesterol, respectively.
The top phase diagram corresponds to higher temperatures, while the bottom one 
to lower temperatures.
The solid lines denote coexistence curves, thin lines are tie lines, and the
numbers are the values of the $\delta$ parameter representing the orientational order
of the saturated lipids.
Adapted from Ref.~\cite{PS08}.
}
\label{fig:putzel}
\end{center}
\end{figure}

The $J_{\rm us}$ term indicates that the effective repulsive interaction
between unsaturated and saturated lipids becomes stronger when the orientational
order $\delta$ increases.
On the other hand, the negative term, proportional to ${-}J_{\rm cs} \delta$, expresses the tendency of cholesterol to increase the chain order $\delta$ of saturated lipids. 
Since the total repulsive interaction between  unsaturated and  saturated
lipids is more enhanced according to this combined effect of cholesterol,
the two lipids tend to segregate when cholesterol is present.
The phase diagram, obtained by minimizing the above free-energy, is presented in
Fig.~\ref{fig:putzel} (top), and qualitatively reproduces the experimental
closed-loop diagram, as shown in Fig.~\ref{fig:ternaryphase} (top).
Moreover, by considering the free energy of the gel phase at lower temperatures,
the theoretical phase diagram Fig.~\ref{fig:putzel} (bottom) qualitatively reproduces
the experimental one, shown in Fig.~\ref{fig:ternaryphase} (bottom).

%%%%%%%%%%%%%%%%%%%%%%%%%%%%%%%%%%%%%%%%%%%%%%%%%%%%%%%%%%%%%%%%%%%%%%%%%
\subsection{Coupling between the two membrane leaflets}
%%%%%%%%%%%%%%%%%%%%%%%%%%%%%%%%%%%%%%%%%%%%%%%%%%%%%%%%%%%%%%%%%%%%%%%%%

In biomembranes of living cells, the two monolayers (leaflets)
have in general different composition, with a unique asymmetry between the inner and outer leaflets~\cite{AZA}.
This asymmetry is essential to the biological function of the cell and is
maintained by active processes such as lipid {\it "flip-flop"}.
Furthermore, the two leaflets are not independent, but rather interact strongly
with each other due to various physical and chemical
mechanisms~\cite{collins-08,May}.

One of the interesting consequences is that a phase separation occurring in one leaflet can affect
the other leaflet in a complex way.
Using the Montal--M\"uller technique, Collins \textit{et al.}~\cite{CK}
addressed the leaflet asymmetry within an artificially
constructed bilayer, which biomimics the {\it in-vivo} situation.
Two lipid monolayers have been combined, after each of them being individually prepared as a Langmuir
monolayer with its own lipid composition, and the phase behavior
was investigated for such coupled asymmetric monolayers.
When one monolayer having a composition that does not exhibit phase separation
was coupled with a second monolayer that was in its two-phase coexistence state,
a phase separation was induced in the former monolayer.
In addition, the experiment has shown strong positional
correlation and domain registration between domains across the two membrane leaflets.

Inspired by the above experiment, a few  phenomenological
models have been proposed~\cite{WLM,PS} to describe the  phase separation in such coupled 
leaflets.
Several suggestions have been made about the possible physical origin
of leaflet coupling, and they include van der Waals
interactions, electrostatic interactions, or mutual interdigitation of
hydrocarbon chains~\cite{May}.

%%%%%%%%%%%%%%%%%%%%%%%%%%%%%%%%%%%%%%%%%%%%%%%%%%%%%%%%%%%%%%%%%%%%%%%%%
\section{Phase separation and conformation changes in membranes}
\label{sec:vesicle}
%%%%%%%%%%%%%%%%%%%%%%%%%%%%%%%%%%%%%%%%%%%%%%%%%%%%%%%%%%%%%%%%%%%%%%%%%
\subsection{Domain-induced budding}
%%%%%%%%%%%%%%%%%%%%%%%%%%%%%%%%%%%%%%%%%%%%%%%%%%%%%%%%%%%%%%%%%%%%%%%%%

Two-dimensional lipid bilayers can take various conformations in the three-dimensional (3d) embedded space.
The observation that even vesicles consisting of a single component lipid exhibit
a variety of complex shapes has been known for a long time.
The most well-known model that describes the shapes of lipid membranes or
vesicles is the {\it "spontaneous curvature model"} pioneered by Helfrich in the early
1970's~\cite{helfrich}.
According to Helfrich model, a membrane is represented as a 2d curved surface of zero thickness,
and its shape is governed by the following curvature elasticity free-energy
\begin{equation}
F_{\rm c}=\frac{\kappa}{2} \int {\rm d} A \, (C_1 + C_2 -2C_0)^2
+ \bar{\kappa} \int {\rm d} A\, C_1 C_2.
\label{curvature}
\end{equation}
In the above equation, $C_1$ and $C_2$ are the two principle curvatures of the
membrane surface related to the mean  and  Gaussian curvatures,
$C=(C_1+C_2)/2$ and $G=C_1C_2$, respectively,
and the
integration is performed over the entire surface area $A$ of the membrane.
The coefficients $\kappa$ and $\bar{\kappa}$ are the bending  and
saddle-splay moduli, respectively, $C_0$ is the material-dependent spontaneous
curvature, reflecting any potential asymmetry between the two sides of the membrane.
For a vesicular shape, by minimizing Eq.~(\ref{curvature}) under the condition
that both the total area $A$ and total inner volume $V$ are conserved, one
can obtain a variety of vesicle shapes that are mechanically
stable\footnote{There is a variety of related models such as
the {\it "bilayer coupling model"}~\cite{SZ} and the {\it "area difference elasticity
model"}~\cite{MSWD}, which also describe vesicular shapes.
For a comparison between these different models, the reader is referred to
Ref.~\cite{lipowsky-sackmann}.}.

What happens to the membrane shape when the lateral phase separation takes
place in a multi-component lipid bilayer?
The line tension acting at the edge of the 2d domains due to
the phase separation plays here an important role.
Prior to any experimental works, Lipowsky~\cite{Lipowsky92}  proposed the idea of
domain-induced budding in multi-component membranes, to be reviewed next.

\begin{figure}[t]
\begin{center}
\resizebox{0.9\columnwidth}{!}{
\includegraphics{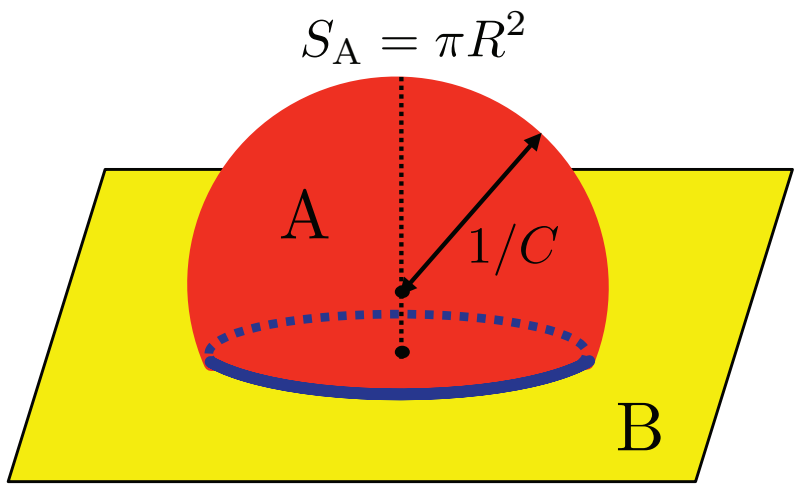}
}
\caption{
A bud (A-domain) forming a spherical cap of radius $1/C$ where
$C$ is the curvature, embedded in a flat B-domain.
The total area of the A-domain is $S_{\rm A}=\pi R^2$.
The line tension $\sigma$ is acting along the boundary
(blue line) between the A and B domains.
}
\label{fig:bud}
\end{center}
\end{figure}

Let's us consider a single 2d circular domain
of radius $R$ embedded in a flat 2d membrane, as shown in Fig.~\ref{fig:bud}.
The domain is characterized by a line tension $\sigma$ that acts
at the 1d edge (perimeter)  of the domain.
Budding of the domain is a process where the domain protrudes in the third dimension (perpendicular to the membrane plane).
For simplicity, we assume that the budded domain shape is
a spherical section of a sphere of radius $1/C$.
The total energy of the budded domain is given by the sum of the curvature
elasticity energy, Eq.~(\ref{curvature}), and the line energy that is
proportional to the domain boundary length (the {\it "neck"}):
\begin{equation}
F_{\rm d}=2\pi \kappa
\left[ (RC-RC_0)^2 + (R/\ell)\sqrt{1-(RC/2)^2} \right],
\label{domain}
\end{equation}
where $\ell=\kappa/\sigma$ is called the
{\it "invagination length"}, and $RC$ is the dimensionless curvature.

The boundary values $RC=\pm 2$ correspond to the {\it "complete budding"},
while $RC=0$ represents the 2d flat domain.
When the spontaneous curvature is nonzero ($C_0 \neq 0$), the symmetry between
the two sides of the membrane is broken.
If the value of $R/\ell$ is small enough, the minimization of total domain energy,
Eq.~(\ref{domain}), yields intermediate $RC$ values,
$0< \vert RC \vert <2$.
This situation for which the
curvature elasticity energy and the line energy balances
each other is called {\it "incomplete budding"}.

%%%%%%%%%%%%%%%%%%%%%%%%%%%%%%%%%%%%%%%%%%%%%%%%%%%%%%%%%%%%%%%%%%%%%%%%%
\subsection{Phase separation in multi-component vesicles}
%%%%%%%%%%%%%%%%%%%%%%%%%%%%%%%%%%%%%%%%%%%%%%%%%%%%%%%%%%%%%%%%%%%%%%%%%

\begin{figure}[t]
\begin{center}
\resizebox{0.9\columnwidth}{!}{
\includegraphics{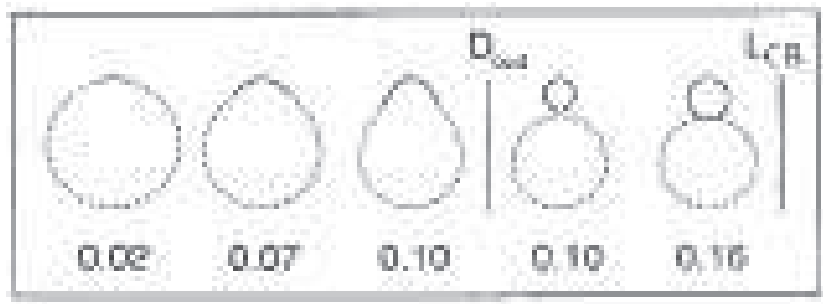}
}
\caption{
Equilibrium shapes of vesicles consisting of two domains (A-domain: solid line,
B-domain: dashed line).
All these shapes are axisymmetric.
The numbers indicate the areal fraction of the A-domain, $\phi_{\rm A}$.
A discontinuous budding transition occurs at D$_{\rm bud}$, while
a singular limit shape with closed neck occurs at L$_{\rm CB}$.
Adapted from Ref.~\cite{JL96}.
}
\label{fig:julicher}
\end{center}
\end{figure}

In Lipowsky's model, the embedding matrix surrounding the domain was assumed to
be infinitely large and flat.
Later, domain-induced budding for closed-shaped and curved vesicles was investigated in great
detail by J\"ulicher and Lipowsky~\cite{JL93,JL96}.
Assuming that a vesicle is composed of two coexisting domains of type A and B,
J\"ulicher and Lipowsky considered
the curvature elasticity energy, Eq.~(\ref{curvature}), for the two domains,
together with the line tension $\sigma$ acting at the domain boundary.
The total free energy
was minimized under the constraint of constant area and volume of the vesicle.
The obtained equilibrium vesicle shape~\cite{JL96} as function of
the $\phi_{\rm A}$, the area fraction of the A-domain (represented by a solid line) is shown 
in Fig.~\ref{fig:julicher}.
For $\phi_{\rm A}=0.1$, a discontinuous transition from incomplete budding
to complete budding takes place.
When $\phi_{\rm A}=0.16$, each domain forms a sphere on its own, and the neck
connecting the two spheres disappears.

\begin{figure}[t]
\begin{center}
\resizebox{1.0\columnwidth}{!}{
\includegraphics{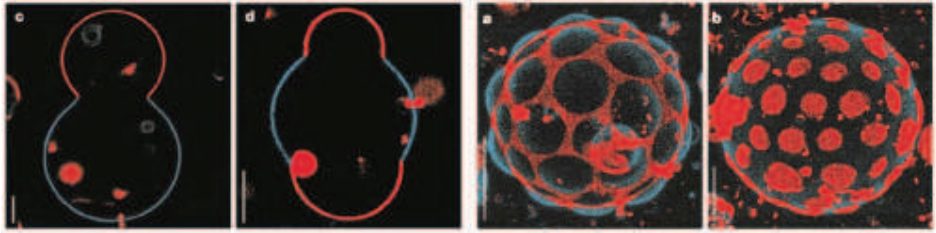}
}
\caption{
Two-photon microscopy images of giant vesicles consisting of DOPC/SM/ cholesterol.
Blue and red domains correspond to the L$_{\rm o}$-phase and the L$_{\rm d}$-phase,
respectively.
In the left two pictures, different phases have different curvatures.
In the right two pictures, finite-sized domains are ordered in a  periodical fashion.
The scale bar corresponds to 5 $\mu$m.
Adapted from Ref.~\cite{BHW}.
}
\label{fig:baumgart}
\end{center}
\end{figure}

Baumgart \textit{et al.}~\cite{BHW} considered experimentally the interplay
between the shape and lateral phase separation in vesicles.
We show the results in Fig.~\ref{fig:baumgart} (left), for vesicles composed of
DOPC/SM\footnote{SM: sphingomyelin (saturated lipid, one type of
sphingolipid)}/ cholesterol. These results demonstrate that each domain is characterized
by a distinct curvature, and multi-domain vesicles form spontaneously complex
structures.
The analysis of these vesicle shapes showed that the observed morphology
can be explained in terms of the model of J\"ulicher and Lipowsky~\cite{BDWJ}, although 
 some of the experimental reported shapes are probably only metastable.

Moreover, Baumgart \textit{et al.}~\cite{BHW} sometimes found vesicles that do not
undergo macroscopic phase separation, but rather form 2d ordered patterns of
finite-size domains as is reproduced in Fig.~\ref{fig:baumgart} (right).
Such patterns are an indication of the so-called
{\it micro-phase separation} --- a well-studied phenomena 
characterizing equilibrium structures of block copolymers
and surfactant solutions~\cite{hamley}.
Although it still remains unclear under what conditions micro-phase separation
can be obtained in bilayers, some recent studies~\cite{KGAHWF,YSIY} have reported that
micro-phase separation can be induced by controlling the composition of membranes composed of
four-component lipid mixtures.
The appearance of such micro-phase separated structures in membranes or vesicles
has been explained in terms of the curvature instability
mechanism~\cite{LA,AKK,KAKT,TKAK}.
Several possible physical mechanisms, including the curvature instability that
leads to micro-phase separations will be addressed separately
in Sec.~\ref{sec:microemulsion} below.

Yanagisawa \textit{et al.}~\cite{YIT} found that when salt was added to multi-component
vesicles in order to control the osmotic pressure difference, complex shape
transformations took place followed by domain budding.
Furthermore, multi-component membranes containing charged lipids have been
also studied experimentally~\cite{Shimokawa}.
In general, phase separation is suppressed because it is energetically
unfavorable to form charged domains~\cite{SKA}.

%%%%%%%%%%%%%%%%%%%%%%%%%%%%%%%%%%%%%%%%%%%%%%%%%%%%%%%%%%%%%%%%%%%%%%%%%
\section{Hydrodynamic effects in fluid membranes}
\label{sec:hydro}

\subsection{Mobility tensor of fluid membranes}
%%%%%%%%%%%%%%%%%%%%%%%%%%%%%%%%%%%%%%%%%%%%%%%%%%%%%%%%%%%%%%%%%%%%%%%%%

So far, we have regarded the equilibrium properties of multi-component
membranes and vesicles. We proceed by reviewing
the hydrodynamic properties of fluid membranes and discuss, in particular, the
dynamics of their heterogeneities.
In their pioneering work, Saffman and Delbr\"uck considered a hydrodynamic model for
biomembranes based on the fluid mosaic model~\cite{saffman-75,saffman-76}.
Their main purpose was to obtain the diffusion coefficient of a protein
molecule embedded inside a fluid membrane.
Since the lateral extent of the membrane surface is typically much larger
than its thickness, it is justified to regard the membrane as a 2d fluid of
zero thickness.
However, if one employs the Stokes approximation in order to analyze the
stationary motion of an isolated object embedded in an infinitely extended 2d fluid, we are
faced with the so-called Stokes paradox~\cite{landaulifshitz-flumech}, where 
the hydrodynamic equation does not have a solution.
The Stokes paradox originates from the inability to conserve momentum within a
2d fluid.

On the other hand, because a fluid membrane is not an isolated 2d
system but surrounded by a 3d solvent (water), the Stokes paradox
can be avoided if 
the dissipation of the membrane momentum is included 
into the surrounding 3d solvent.
The obtained diffusion coefficient is given below
by Eq.~(\ref{eqn:longDsmallr}), and has been extensively used in analyzing
experimental data~\cite{lipowsky-sackmann}.
Hereafter, we consider a slightly more generalized situation and explain  how to
derive the hydrodynamic mobility tensor.

\begin{figure}[t]
\begin{center}
\resizebox{0.7\columnwidth}{!}{
\includegraphics{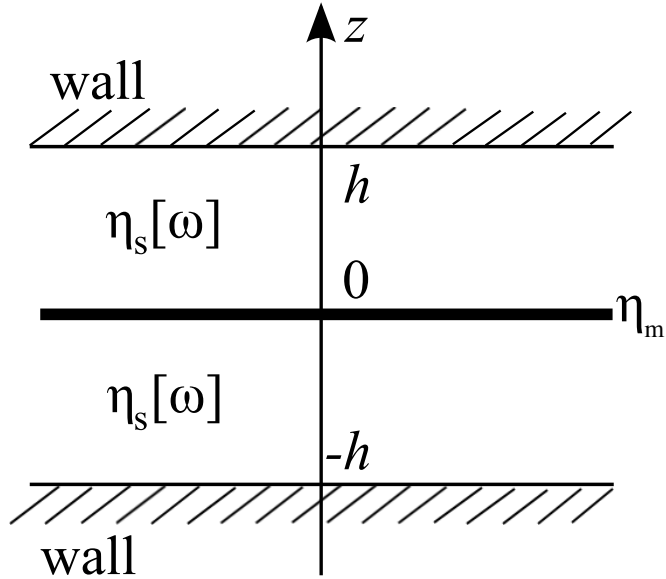}
}
\caption{
Generalized Saffman and Delbr\"uck hydrodynamic model.
Planar viscous membrane at $z=0$ of 2d viscosity $\eta_{\rm m}$ is sandwiched in between
two solvents having the same 3d viscosity $\eta_{\rm s}$.
Two impenetrable walls at $z=\pm h$ bound the upper and lower solvents.
When the solvent is viscoelastic, its viscosity $\eta_{\rm s}[\omega]$ becomes
frequency dependent.
}
\label{fig:hydro}
\end{center}
\end{figure}

Consider an infinitely extended 2d  flat fluid membrane
of 2d viscosity $\eta_{\rm m}$, as shown in
Fig.~\ref{fig:hydro}~\cite{sanoop-poly-10,SRK11},
sandwiched between an upper and a lower 3d solvents of the same 3d viscosity
$\eta_{\rm s}$. Note that the units of the 3d viscosity $\eta_{\rm s}$ and 2d
viscosity $\eta_{\rm m}$ are different. Two solid walls are placed at distance $h$
from the membrane making the thickness of the solvents finite.
This setup is motivated by many experiments that are
performed on supported lipid bilayers placed on top of a solid
substrate\footnote{The result is almost the same even if there is only
one wall.}.
The inplane velocity vector of the fluid membrane is denoted by
${\mathbf v}({\mathbf r})$ where ${\mathbf r}=(x,y)$ is a 2d position vector.
Assuming that the incompressibility condition holds for the fluid membrane,
we write its hydrodynamic equations as
\begin{equation}
\nabla \cdot {\mathbf v} = 0,\nonumber
\end{equation}
\begin{equation}
\eta_{\rm m} \nabla^2 {\mathbf v} - \nabla p + {\mathbf f}_{\rm s} + {\mathbf F}=0.
\end{equation}
The second equation is the 2d Stokes equation,
where $p$ is the lateral pressure, ${\mathbf f}_{\rm s}$ is the force exerted
on the membrane by the surrounding solvent, and ${\mathbf F}$ is any external force
acting on the membrane.

If we denote the upper and lower solvents with the superscripts $\pm$,
the two solvent velocities ${\mathbf v}^{\pm}({\mathbf r},z)$ and
pressures $p^{\pm}({\mathbf r},z)$ obey the following hydrodynamic equations,
respectively
\begin{equation}
\widehat{\nabla} \cdot {\mathbf v}^{\pm} = 0, \nonumber
\end{equation}
\begin{equation}\eta_{\rm s} \widehat{\nabla}^2 {\mathbf v}^{\pm} - \widehat{\nabla} p^{\pm}= 0,
\end{equation}
where $\widehat{\nabla}$ stands for the 3d differential operator (while for convenience $\nabla$ denotes 
the operator in 2d).

The boundary conditions at the membrane are non-slip leading to matching of the membrane and solvent velocities
at the membrane plane, $z=0$. Furthermore, the solvent velocity vanishes at the $z=\pm h$ walls.
With the use of the solvent stress tensor we can write
\begin{equation}
{\bm \sigma}^{\pm} = - p^{\pm} {\mathbf I}
+ \eta_{\rm s} [\widehat{\nabla} {\mathbf v}^{\pm}
+ (\widehat{\nabla} {\mathbf v}^{\pm})^{\rm T}],
\end{equation}
where the 'T' superscript stands for the transpose operator,
the force ${\mathbf f}_{\rm s}$ is given by the projection of
$\hat{{\mathbf e}}_z \cdot ({\bm \sigma}^{+} - {\bm \sigma}^{-})_{z=0}$ onto the
membrane, $\hat{{\mathbf e}}_z$ is the unit vector in the $z$-direction.
By solving the above coupled hydrodynamic equations in Fourier space with
$\mathbf{k}=(k_x,k_y)$ being the 2d wavevector, the 2d mobility
tensor ${G}_{ij}({\mathbf k})$ defined through
${ v_i}({\mathbf k})= \sum_j{ G}_{ij}({\mathbf k})  { F}_{j}({\mathbf k})$
($i,j=x,y$) can now be written as
\begin{equation}
G_{ij}({\mathbf k}) =  \frac{1}{\eta_{\rm m} [ k^2 + \nu k \coth(kh)]}
\left( \delta_{ij} - \frac{k_i k_j}{k^2} \right),
\label{eqn:cothoseen}
\end{equation}
where $\nu^{-1} = \eta_{\rm m}/2 \eta_{\rm s}$ is the hydrodynamic screening
length and $k=\vert {\mathbf k} \vert$.
We note that the above  membrane mobility tensor is analogous to the Oseen tensor for 3d
fluids~\cite{landaulifshitz-flumech}, and
the presence of the surrounding solvent of finite thickness (solid walls at $z=\pm h$)
is accounted for by the second term of the denominator in Eq.~(\ref{eqn:cothoseen}).

%%%%%%%%%%%%%%%%%%%%%%%%%%%%%%%%%%%%%%%%%%%%%%%%%%%%%%%%%%%%%%%%%%%%%%%%%
\subsection{Coupling diffusion: free membrane case}
%%%%%%%%%%%%%%%%%%%%%%%%%%%%%%%%%%%%%%%%%%%%%%%%%%%%%%%%%%%%%%%%%%%%%%%%%

Saffman and Delbr\"uck considered the limiting case of $kh\gg1$ in
Eq.~(\ref{eqn:cothoseen}), for which the denominator can be approximated by
$\eta_{\rm m} ( k^2 + \nu k)$.
This situation is equal to the free membrane case because there is
no dependence on the bounding walls and solvent thickness, $h$.
We consider two point-particles (particle 1 and 2) separated by distance
$r$ on the membrane, and discuss the longitudinal coupling diffusion
coefficient $D_{\rm L}$
defined by $\langle \Delta x_1 \Delta x_2 \rangle_r=2 D_{\rm L}(r) t$,
where $\Delta x_i$ is the displacement of the $i$-th particle along the line
connecting the two point-particles, and
$t$ is time\footnote{The
$x$-axis is chosen to be parallel to the line connecting the two point-particles.
One can also consider the correlation of the displacements along the
$y$-axis perpendicular to the line connecting the two particles. 
This gives the transverse coupling
diffusion coefficient, $D_{\rm T}(r)$.}.
By taking the inverse Fourier transform of the mobility tensor and
using the Einstein relation, the coupling diffusion coefficient is obtained
as~\cite{oppenheimer-09,oppenheimer-10}

\begin{equation}
D_{\rm L}(r)=\frac{k_{\rm B}T}{4\pi \eta_{\rm m}}
\left[\frac{\pi {\mathbf H}_1(\nu r)}{\nu r}-
\frac{\pi Y_1(\nu r)}{\nu r}
- \frac{2}{(\nu r)^2} \right],
\label{eqn:longD}
\end{equation}
where ${\mathbf H}_1(x)$ and $Y_1(x)$ are the Struve function and the
Bessel function of the second kind, respectively.

There are two asymptotic limits of Eq.~(\ref{eqn:longD}) depending on the value of $\nu r$.
In the small separation limit, $\nu r \ll 1$,

\begin{equation}
D_{\rm L}(r) \approx
\frac{k_{\rm B}T}{4 \pi \eta_{\rm m}}
\left[ \ln\left(\frac{2}{\nu r}\right) - \gamma + \frac{1}{2}\right],
\label{eqn:longDsmallr}
\end{equation}
where $\gamma=0.5772 \cdots$ is the Euler's constant, whereas in the
large separation limit, $\nu r \gg 1$

\begin{equation}
D_{\rm L}(r) \approx
\frac{k_{\rm B}T}{2\pi\eta_{\rm m}\nu r}=
\frac{k_{\rm B}T}{4\pi\eta_{\rm s} r}.
\label{eqn:longDlarger}
\end{equation}
It is worth mentioning that the coupling diffusion coefficient
$D_{\rm L}(r)$ between  two point-particles separated by a distance $r$ corresponds
to the self-diffusion coefficient of a single particle of size $r$
up to a prefactor of order unity.
			
The above results clearly demonstrate the hydrodynamic behavior of a 2d
fluid membrane surrounded by 3d solvent.
If the distance $r$ between two points on the membrane is sufficiently small compared
with the hydrodynamic screening length $\nu^{-1}$, the diffusion coefficient
is almost independent of $r$
(see Eq.~(\ref{eqn:longDsmallr}))~\cite{saffman-75,saffman-76}.
On the other hand, if $r$ is sufficiently large as compared with $\nu^{-1}$, $D_{\rm L}$
is inversely proportional to $r$ (see Eq.~(\ref{eqn:longDlarger})), 
similar to the Stokes--Einstein relation for a 3d solid
sphere~\cite{hughes-81}.

In other words, Eq.~(\ref{eqn:longDsmallr}) reflects the 2d nature of the
fluid membrane, while Eq.~(\ref{eqn:longDlarger}) represents the 3d nature
of the outer solvent.
Since typical values of the hydrodynamic screening length $\nu^{-1}$ are
in the sub-micron range, $\nu r \ll 1$ holds for usual membrane
proteins whose size is in the nanometer rage.
In contrast, it was experimentally demonstrated that the diffusion
coefficient of micron-sized domains (much larger than a protein molecule),
is inversely proportional to their size~\cite{cicuta-07}, in agreement with Eq.~(\ref{eqn:longDlarger}).

%%%%%%%%%%%%%%%%%%%%%%%%%%%%%%%%%%%%%%%%%%%%%%%%%%%%%%%%%%%%%%%%%%%%%%%%%
\subsection{Coupling diffusion: confined membrane case}
%%%%%%%%%%%%%%%%%%%%%%%%%%%%%%%%%%%%%%%%%%%%%%%%%%%%%%%%%%%%%%%%%%%%%%

The other limit of $kh\ll1$ in Eq.~(\ref{eqn:cothoseen}) corresponds to the
confined membrane case, where the lipid bilayer is supported by
a solid substrate (if there is only one wall).
Evans and Sackmann were the first to consider such a case~\cite{evans-88},
and Seki and Komura and coworkers~\cite{seki-93,komura-95,sanoop-drag-10}
applied it to membranes, whose
momentum dissipates into the surrounding solvent with a characteristic decay rate.

For confined membranes the denominator in Eq.~(\ref{eqn:cothoseen})
can be approximated by $\eta_{\rm m}(k^2+\chi^2)$~\cite{stone-98}, where 
$\chi^{-1} = \sqrt{\nu^{-1}h}$ is the hydrodynamic screening length.
The corresponding coupling
diffusion coefficient can be obtained similarly to Eq.~(\ref{eqn:longD}), 
and results in a logarithmic
dependence when $\chi r \ll 1$.
For $\chi r \gg 1$, however, $D_{\rm L}(r) \sim h/r^2$, which decays as
$\sim 1/r^2$ rather than $\sim 1/r$ as in Eq.~(\ref{eqn:longDlarger}).

%%%%%%%%%%%%%%%%%%%%%%%%%%%%%%%%%%%%%%%%%%%%%%%%%%%%%%%%%%%%%%%%%%%%%%%%%
\subsection{Effects of solvent viscoelasticity}
%%%%%%%%%%%%%%%%%%%%%%%%%%%%%%%%%%%%%%%%%%%%%%%%%%%%%%%%%%%%%%%%%%%%%%%%%

In eukaryotic cells, the cytoplasm contains proteins, subcellular
organelles as well as an actin meshwork forming the cell cytoskeleton~\cite{alberts}.
In addition, the outside of the cell is composed of an extracellular matrix
and/or hyaluronic acid gel that can be regarded as a polymer solution.

Komura \textit{et al.}~\cite{KRS12,KRS12-2} discussed the dynamics of biomembranes under the
assumption that the surrounding solvent is viscoelastic, and
we mention it here.
The surrounding solvent was considered to obey the
constitutive equation:
\begin{equation}
{\bm \sigma}^{\pm}(t)= 2 \int_{-\infty}^{t} {\rm d}t'
\eta_{\rm s}(t - t') \mathbf{D}^{\pm}(t'),
\end{equation}
where $\eta_{\rm s}(t)$ is the time-dependent solvent viscosity, and
\begin{equation}
\mathbf{D}^{\pm}=\frac{1}{2}
[ \widehat{\nabla} \mathbf{v}^{\pm} +
(\widehat{\nabla} \mathbf{v}^{\pm})^{\rm T} ],
\end{equation}
is the rate-of-strain tensor.

By repeating the calculation along the lines done in Sec.~4.1, we obtain the mobility
tensor $G_{ij}(\mathbf{k},\omega)$ that  depends also on the
frequency $\omega$.
For simplicity, let us assume that the frequency dependence
of the solvent viscosity obeys a power law:
$\eta_{\rm s}(\omega) = G_0 (i \omega)^{\beta-1}$ with $\beta<1$
because $\eta_{\rm s}(\omega)$ should vanish for
$\omega \rightarrow \infty$.
Notice that the purely viscous case is recovered for $\beta \rightarrow 1$.
Using the fluctuation-dissipation theorem, it is possible to calculate
the time-dependence of the two-particle correlation function
$\langle \Delta x_1 \Delta x_2 \rangle_r$ from $G_{ij}(\mathbf{k},\omega)$.
For large $r$, the correlation behaves asymptotically as
$\langle \Delta x_1 \Delta x_2\rangle_r \sim
(k_{\rm B}T/G_0 r) t^{\beta}$, which gives rise to anomalous
diffusion since $\beta<1$~\cite{KRS12,KRS12-2}.
Recently, anomalous diffusion of membrane proteins has been experimentally
observed~\cite{weigel}, but it should be equally
noted that there are other mechanisms that may lead
to such anomalous diffusion.

%%%%%%%%%%%%%%%%%%%%%%%%%%%%%%%%%%%%%%%%%%%%%%%%%%%%%%%%%%%%%%%%%%%%%%%%%
\section{Dynamics of lateral phase separation}
\label{sec:separation}

\subsection{Experiments on domain growth-law}
%%%%%%%%%%%%%%%%%%%%%%%%%%%%%%%%%%%%%%%%%%%%%%%%%%%%%%%%%%%%%%%%%%%%%%%%%

\begin{figure}[t]
\begin{center}
\resizebox{1.0\columnwidth}{!}{
\includegraphics{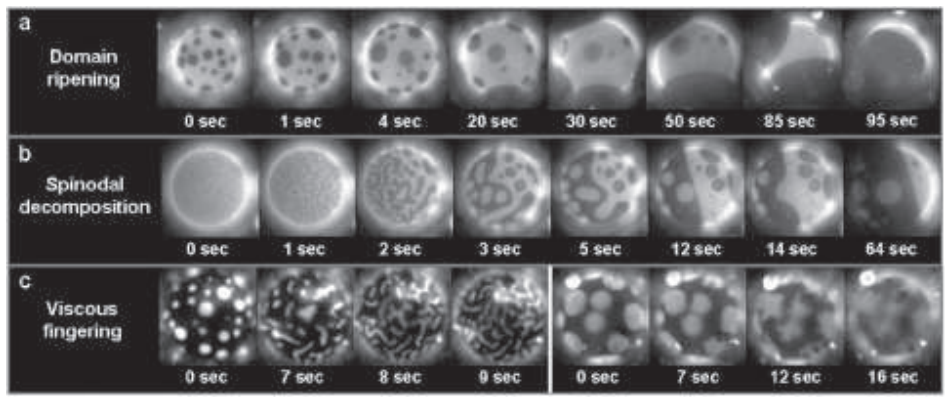}
}
\caption{
Time evolution of phase separation observed for ternary vesicles consisting
of DOPC/DPPC/ cholesterol.
(a) Domain growth (ripening) due to collision and coalescence when the ratio of the
L$_{\rm o}$-phase and L$_{\rm d}$-phase is asymmetric.
(b) Spinodal decomposition when the two phases are symmetric.
(c) Viscous fingering when the two phases are highly asymmetric.
Adapted from Ref.~\cite{veatch-03}.
}
\label{fig:dynamics}
\end{center}
\end{figure}

Investigations on phase separation in multi-component lipid bilayers initially
focused on equilibrium properties such as phase behavior, as was explained
in the previous sections.
Later, substantial attention has been devoted to the dynamics of phase separation in
membranes.
For ternary mixtures of DOPC/DPPC/ cholesterol, Veatch \textit{et al.}~\cite{veatch-03} reported
several types of growth patterns depending on the relative
membrane composition.
When the area fraction between the L$_{\rm o}$-phase and  L$_{\rm d}$-phase is
asymmetric, as shown in Fig.~\ref{fig:dynamics}(a), domain growth occurs.
The process is dominated by the collision-coalescence mechanism rather than by
the evaporation-condensation mechanism.
When the area ratio between the two phases was almost symmetric, namely
1:1, spinodal decomposition was observed, as shown in Fig.~\ref{fig:dynamics}(b).

When a dynamical scaling law holds, the average domain size $R$ increases
according to a temporal power-law,
$R(t) \sim t^{\alpha}$.
Using fluorescence microscope, Saeki \textit{et al.}~\cite{saeki-06} performed a quantitative
measurement of the growth exponent $\alpha$ for vesicles consisting of
DOPC/DPPC/ cholesterol, and reported the value
$\alpha \approx 0.15$.
Later, Yanagisawa \textit{et al.}~\cite{yanagisawa-07} conducted a similar experiment and found
that there are two different types of domain growth that depend on system conditions (explained below).
The first type is due to the collision-coalescence mechanism, with a 
growth exponent,  $\alpha \approx 2/3$.
For the second type, the domain growth was suppressed over a long period of
time, although the domain size suddenly increased at the final stage of the
phase separation. 

Even though the conditions to distinguish between the two types of domain growth 
is not completely clear,
the collision-coalescence mechanism is more dominant when the excess area
of the vesicle is relatively
small\footnote{The excess area is defined by $R_A/R_V-1$ where $R_A$ and
$R_V$ are, respectively, the radii of  spheres having identical area
and volume of the corresponding vesicle.}.
When the excess area is large, budding of domains can take place, and the
elastic interaction between the domains mediated by the membrane affects
the phase separation dynamics. Finally, we note that
recently Stanich \textit{et al.}~\cite{Stanich} performed a systematic experimental study
on the growth exponent.
They reported $\alpha=0.29 \pm 0.05$ for asymmetric compositions, and
$\alpha=0.31 \pm 0.05$ for nearly symmetric compositions when the
collision-coalescence mechanism was dominant.

%%%%%%%%%%%%%%%%%%%%%%%%%%%%%%%%%%%%%%%%%%%%%%%%%%%%%%%%%%%%%%%%%%%%%%%%%
\subsection{Collision-coalescence mechanism of domain growth}
%%%%%%%%%%%%%%%%%%%%%%%%%%%%%%%%%%%%%%%%%%%%%%%%%%%%%%%%%%%%%%%%%%%%%%%%%

Considerable theoretical interest has been devoted to the understanding
of dynamics of phase separation in lipid membranes, and, in particular,
several numerical studies were conducted~\cite{Taniguchi,SGL,AMMV} using computer
simulations.
Laradji \textit{et al.}~\cite{degroot-warren-97} used dissipative 
particle-dynamics method in order to simulate a two-component vesicle
composed of coarse-grained lipid molecules~\cite{LK04,sunil-mohamed-05}.
When the composition of the two lipids was asymmetric, the domain growth was
driven by the collision-coalescence mechanism, and the growth exponent was
found to be $\alpha \approx 0.3$.
Moreover, it was reported that even budding of domains occurs when the
excess area was large enough.

\begin{figure}[t]
\begin{center}
\resizebox{0.8\columnwidth}{!}{
\includegraphics{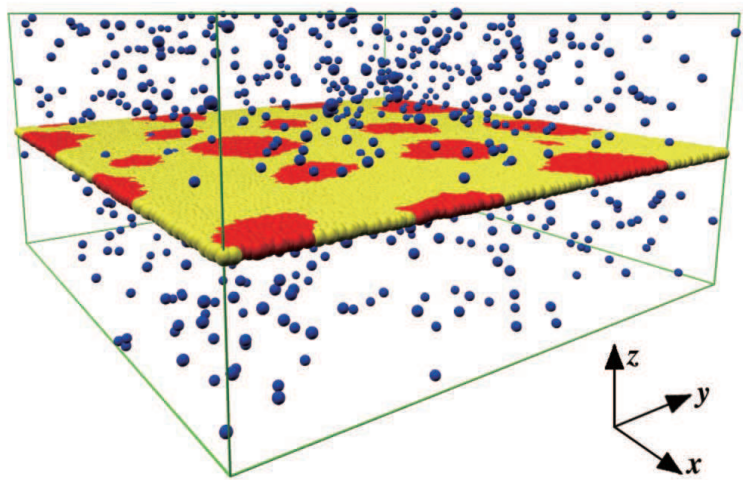}
}
\caption{
Binary fluid membrane with the surrounding solvent.
The yellow (lipid A) and red (lipid B) particles represent the two components
constituting the membrane, while the blue particles represent the solvent.
For clarity, only a fraction of the solvent particles are shown.
}
\label{fig:ramachandran}
\end{center}
\end{figure}

Ramachandran \textit{et al.}~\cite{RKG10} performed a dissipative particle-dynamics simulation
in order to investigate the hydrodynamic effects on the 2d membrane phase-separation
when the membrane is placed in contact with a 3d solvent.
As shown in Fig.~\ref{fig:ramachandran}, it was assumed that a flat fluid membrane
is composed of A (yellow) and B (red) particles. The membrane is sandwiched
by 3d solvent  particles (blue), and its bilayer nature is neglected.
The time evolution of the phase separation between the pure-2d membrane
(no solvent) was compared with that of a quasi-2d one (with solvent particles).
The result showed that the domain size in the quasi-2d case was smaller than
for the pure-2d case over the same period of time. It offers an evidence that the
phase separation in the quasi-2d membrane is suppressed.
Even more quantitative analysis~\cite{RKG10} reveals that the growth exponent for the
pure-2d case is $\alpha=1/2$, while for quasi-2d case it is
$\alpha=1/3$.

The various values of the  growth exponent $\alpha$ can be explained as follows.
The domain growth occurs through the collision-coalescence process
driven by the Brownian motion of domains.
If the domain size $R$ is the only relevant length scale, the scaling relation
$R^2 \sim D t$ should hold, where $D$ is the domain diffusion coefficient,
as discussed in the previous section.
Since the hydrodynamic screening length $\nu^{-1}$ for the pure-2d membrane is
considered to be infinitely large, $R$ is always smaller
than $\nu^{-1}$, and the diffusion coefficient $D$ is almost constant\footnote{As mentioned before, the
distance $r$ between the two points corresponds to the domain size $R$.},
as shown in Eq.~(\ref{eqn:longDsmallr}).
Hence, we obtain $\alpha=1/2$ from the scaling $R \sim t^{1/2}$.

For the quasi-2d membrane, on the other hand, $\nu^{-1}$ is finite, and $R$ will
become larger than $\nu^{-1}$ in the late stages of the phase separation.
As a result, the 3d hydrodynamic interactions mediated by the solvent will
become more important, and the diffusion coefficient behaves like
$D \sim 1/R$, as given by Eq.~(\ref{eqn:longDlarger}).
Thus, we obtain $R \sim t^{1/3}$ that explains why the growth exponent for
the quasi-2d membrane is $\alpha=1/3$.
We remark that this value is in accord with the recent experimental
result by Stanich \textit{et al.}~\cite{Stanich}.

%%%%%%%%%%%%%%%%%%%%%%%%%%%%%%%%%%%%%%%%%%%%%%%%%%%%%%%%%%%%%%%%%%%%%%%%%
\subsection{Dynamics of concentration field}
%%%%%%%%%%%%%%%%%%%%%%%%%%%%%%%%%%%%%%%%%%%%%%%%%%%%%%%%%%%%%%%%%%%%%%%%%

In order to describe the dynamics of phase separation in multi-component
biomembranes using a continuous concentration field, one can employ
the time-dependent Ginzburg--Landau (TDGL) model.
The local area fractions (concentrations) of A and B lipids in binary membranes
are denoted by $\phi_{\rm A} ({\mathbf r},t)$ and $\phi_{\rm B} ({\mathbf r},t)$,
respectively.
Since the incompressibility condition is given by $\phi_{\rm A}+ \phi_{\rm B}=1$,
it is enough to introduce an order parameter being the relative concentration:
$\phi= \phi_{\rm A}- \phi_{\rm B}$.
The Ginzburg--Landau free-energy that describes the phase separation of
a binary membrane is given by\footnote{As in Sec.~\ref{sec:hydro},
$\nabla$ is a 2d differential operator, and the integral is also performed in 2d.}

\begin{equation}
F_{\rm GL}[\phi] = \int {\rm d}{\mathbf r} \,
\left[ \frac{a}{2}\phi^2 + \frac{1}{4}\phi^4 +
\frac{c}{2} (\nabla \phi)^2 \right],
\label{eqn:freeE}
\end{equation}
where $a\sim (T-T_{\rm c})$ is the reduced temperature and $c$ is
related to the line tension $\sigma$.

The time evolution of the concentration field $\phi$, which is a conserved
order parameter, can be described by the following dynamical
equation~\cite{chaikin-lubensky}:

\begin{equation}
\frac{\partial \phi}{\partial t} + \nabla \cdot ({\mathbf v}\phi)
= L \nabla^2\frac{\delta F_{\rm GL}}{\delta \phi},
\label{sec:concfluct:eqn4}
\end{equation}
where $L$ is the transport coefficient.
The membrane velocity ${\mathbf v}$ in the second term obeys the hydrodynamic
equations described in Sec.~\ref{sec:hydro}.
The velocity and concentration fields within the 2d membrane are coupled
through the external force given by
$\mathbf{F} = -\phi \nabla (\delta F_{\rm GL}/\delta \phi)$ in the 2d Stokes
equation.
The present model for quasi-2d fluid membranes is an extension of the so-called
{"Model H"} introduced by Hohenberg and Halperin~\cite{hohenberg}
to describe 3d fluids at criticality.

To study the dynamics of phase separation, this extended Model H was
numerically solved by Camley \textit{et al.}~\cite{CB11} and
Fan \textit{et al.}~\cite{fan-10d} in the presence of thermal noise.
In addition to the previously mentioned collision-coalescence mechanism,
domain coarsening in these simulations includes also the evaporation-condensation mechanism,
driven by the line tension $\sigma$ between the domains. It results in a scaling relation
of the domain growth given by
$R \sim (L \sigma t)^{1/3}$~\cite{CB11,lifshitz-81}.
Although the corresponding exponent $\alpha=1/3$ is independent of the spatial
dimension, the growth exponent due to the collision-coalescence mechanism
depends on the relative magnitude of $R$ and $\nu^{-1}$, as mentioned earlier.

When the average composition is asymmetric, various scaling regimes have been
identified in the numerical simulations of the extended Model H~\cite{CB11}.
However, the situation becomes somewhat more complicated when the average composition
is symmetric.
The simulations show that the coarsening of  domains that are isotropic in their shape
is slower than for anisotropic domains.
This means that phase-separated patterns at different times are not
characterized by a single length scale, and indicates a breakdown of the
dynamical scaling law.

%%%%%%%%%%%%%%%%%%%%%%%%%%%%%%%%%%%%%%%%%%%%%%%%%%%%%%%%%%%%%%%%%%%%%%%%%
\subsection{Non-equilibrium effects}
%%%%%%%%%%%%%%%%%%%%%%%%%%%%%%%%%%%%%%%%%%%%%%%%%%%%%%%%%%%%%%%%%%%%%%%%%

It was suggested by several authors~\cite{ST11,Foret12} that raft formation in biomembranes is associated
with the non-equilibrium natures of biomembranes, and potentially includes material exchange
between the membrane and its surroundings.
For example, Foret~\cite{Foret} proposed a time evolution equation for the
concentration field $\phi$ by taking into account the lipid exchange with the
surrounding:
\begin{equation}
\frac{\partial \phi}{\partial t} =
L \nabla^2\frac{\delta F_{\rm GL}}{\delta \phi}
- \Omega (\phi - \bar{\phi}).
\label{foret}
\end{equation}
In the above,  $F_{\rm GL}[\phi]$ is the GL free-energy given by Eq.~(\ref{eqn:freeE}), $\Omega$
the lipid exchange rate, and $\bar{\phi}$ the spatial average value of $\phi$.
It is interesting to note that as Eq.~(\ref{foret}) is formally identical
with the time evolution 
equation of block copolymers, it will lead to a micro-phase
separation of domains, just as is the case in the block copolymer case~\cite{hamley}.
Interested readers are referred to the work by Fan \textit{et al.}~\cite{FSH3} for
other proposed origins of raft formation, such as partitioning effects of lipids,
and the non-equilibrium transport of lipids, as well as to
another model~\cite{GSR09} suggesting cholesterol exchange (instead of lipid exchange)
between the membrane and the outer surrounding solvent.

Recently dynamics of biomembranes mediated by chemical reactions has attracted
some attentions.
Hamada \textit{et al.}~\cite{HSNT} added photoresponsive amphiphile to a typical ternary
lipid mixture, and showed that its conformation change can switch on a reversible lateral
segregation of the membrane.
They demonstrated that {\it cis}-isomerization induces lateral phase separation
in membranes that are in their one-phase (homogeneous) region,
while producing additional lateral domains in membranes that are in their two-phase coexisting regions.

%%%%%%%%%%%%%%%%%%%%%%%%%%%%%%%%%%%%%%%%%%%%%%%%%%%%%%%%%%%%%%%%%%%%%%%%%
\section{Dynamics of concentration fluctuations}
\label{sec:fluctuation}

\subsection{Critical phenomena in membranes}
%%%%%%%%%%%%%%%%%%%%%%%%%%%%%%%%%%%%%%%%%%%%%%%%%%%%%%%%%%%%%%%%%%%%%%%%%

So far we have discussed domain formation that occurs
at temperatures below the phase-separation temperature, and its induced
structural changes in vesicles.
Experimentally, concentration fluctuations above the critical temperature have
been also observed.
Veatch \textit{et al.}~\cite{veatch-07} used {Nuclear Magnetic Resonance} (NMR) to
measure the concentration fluctuations in DOPC/DPPC/ cholesterol mixtures,
and extracted from the data the corresponding correlation-length.

\begin{figure}[t]
\begin{center}
\resizebox{0.9\columnwidth}{!}{
\includegraphics{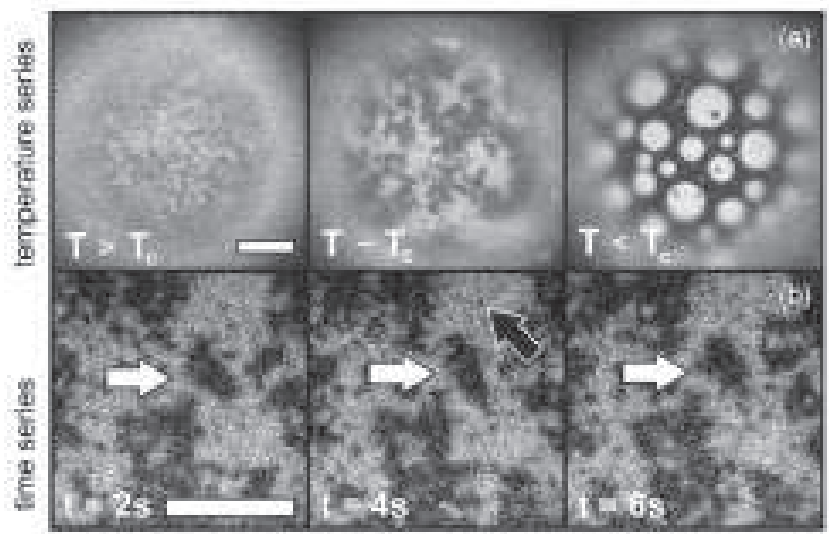}
}
\caption{
Fluorescence microscope images of giant vesicles composed of
diPhyPC/DPPC/ cholesterol.
The critical temperature is $T_{\rm c}=30.9$ $^{\circ}$C.
(a) Concentration fluctuations ($T>T_{\rm c}$, one-phase region), 
criticality at $T\sim T_{\rm c}$, 
and domain growth ($T<T_{\rm c}$, two-phase coexistence region) are observed.
(b) Time evolution of the concentration fluctuations for $T>T_{\rm c}$.
The scale bar corresponds to 20 $\mu$m.
Adapted from Ref.~\cite{hsmith-12}.
}
\label{fig:honerkamp}
\end{center}
\end{figure}

Another ternary mixture (diPhyPC/DPPC/ cholesterol) was used by
Honerkamp-Smith \textit{et al.}~\cite{hsmith-08} to study vesicles
whose membrane composition corresponds to critical phenomena.
We reproduce their fluorescence microscope pictures of concentration fluctuations
in Fig.~\ref{fig:honerkamp}. For these mixtures,
the critical temperature is $T_{\rm c}=30.9$ $^{\circ}$C.
Concentration fluctuations are observed for $T>T_{\rm c}$, while
domain formation are observed for
$T<T_{\rm c}$, (see Fig.~\ref{fig:honerkamp}(a)).
From the experimental data, the critical exponent for the correlation length,
$\xi\sim|T-T_{\rm c}|^{-1.2\pm 0.2}$, was extracted with the conclusion that
the critical behavior of this ternary lipid mixture belongs to the universality
class of the 2d Ising model.

Even more surprising, the 2d Ising critical behavior was also observed in
biomembranes which were extracted from the basophil leukemic cells of a
living rat~\cite{veatch-08}.
These series of experiments proposed the possibility that heterogeneous
structures in biomembranes under physiological conditions can be attributed
to the critical concentration fluctuations.
However, it is rather specific to systems that are in the close proximity
of a critical point, and the application to biomembranes at
physiological conditions is yet to be confirmed.

Recently, Honerkamp-Smith \textit{et al.}~\cite{hsmith-12} have elaborated
also on the dynamics of concentration fluctuations in membranes.
Figure~\ref{fig:honerkamp}(b) shows the time evolution of
the concentration fluctuations and indicates that the structure of the large-scale
fluctuations (white arrow in the figure) is sustained over few seconds,
while the small-scale fluctuations (black arrow in the figure) disappear
almost instantaneously.
The relaxation time of the concentration fluctuations was measured
in order to determine how it increases as the critical point is approached
(critical slowing down).
From the data it was suggested that a dynamic scaling law $\tau \sim \xi^z$ holds between
the relaxation time $\tau$ and the correlation length $\xi$ with an
exponent $z$.
As the critical point is approached from above, the correlation
length grew, and it was found that the
apparent exponent crosses over from $z=2$ to $z=3$.
Next we shall discuss this crossover behavior from a theoretical point
of view.

%%%%%%%%%%%%%%%%%%%%%%%%%%%%%%%%%%%%%%%%%%%%%%%%%%%%%%%%%%%%%%%%%%%%%%%%%
\subsection{Decay rate of concentration fluctuations}
%%%%%%%%%%%%%%%%%%%%%%%%%%%%%%%%%%%%%%%%%%%%%%%%%%%%%%%%%%%%%%%%%%%%%%%%%

The dynamics of concentration fluctuations in biomembranes was modeled by 
Seki \textit{et al.}~\cite{seki-07}.
They used the 2d hydrodynamic model that takes into account the momentum
dissipation (i.e., confined membrane case) to calculate the intermediate
time-dependent structure factor, $S({\mathbf k},t)$.
This quantity is the spatial Fourier transform of the time-dependent
correlation function
$G_{\phi\phi}({\mathbf r},t) =
\langle \delta \phi({\mathbf r},t) \delta \phi(0,0) \rangle$,
where $\delta \phi({\mathbf r},t)$ is the concentration fluctuation.
Solving the equations for the concentration and velocity fields
it was found that the intermediate correlation function decays exponentially:

\begin{equation}
S({\mathbf k},t) = S({\mathbf k},0 ) {\rm e}^{-\Gamma({\mathbf k})t}
\end{equation}
with

\begin{equation}
\Gamma({\mathbf k})=k^2 D_{\rm eff}({\mathbf k}),
\end{equation}
where $\Gamma({\mathbf k})$ is the wavenumber-dependent decay rate (inverse
of the relaxation time), from which the effective
diffusion coefficient $D_{\rm eff}$  can be derived analytically~\cite{seki-07}.
By using the correlation length $\xi=(c/a)^{1/2}$ (see Eq.(\ref{eqn:freeE}))
and the hydrodynamic screening length $\chi^{-1}$ for the confined membrane
case, it was shown, assuming $k \xi \ll 1$, that the asymptotic 
expressions for the diffusion coefficient are
$D_{\rm eff} \sim \ln(1/\xi)$ for $\chi \xi \ll 1$, and
$D_{\rm eff} \sim 1/\xi^2$ for $\chi \xi \gg 1$.

More recently, Inaura \textit{et al.}~\cite{inaura-08} used as their starting point
the free membrane case, and performed a calculation similar
to that of Seki \textit{et al.}\footnote{More precisely,
Seki \textit{et al.}~\cite{seki-07} considered the limit
of $kh \ll 1$ of Eq.~(\ref{eqn:cothoseen}), while Inaura \textit{et al.}~\cite{inaura-08}
considered the opposite limit of $kh \gg 1$.}.
By noting that the hydrodynamic screening length is now given by $\nu^{-1}$,
they confirmed numerically, assuming $k \xi \ll 1$, that the asymptotic behavior is
$D_{\rm eff} \sim \ln(1/\xi)$ for $\nu \xi \ll 1$, and
$D_{\rm eff} \sim 1/\xi$ for $\nu \xi \gg 1$.
When the hydrodynamic interaction is present, the dynamic critical exponent
$z$ can be evaluated by the relation $\tau \sim \xi^2/D_{\rm eff}$.
The result of Inaura \textit{et al.}~\cite{inaura-08} indicates the crossover
behavior of the critical exponent from $z=2$ to $z=3$, which is in agreement
with the experimental findings of Honerkamp-Smith
\textit{et al.}~\cite{hsmith-12}\footnote{Within the confined
membrane case considered by Seki \textit{et al.}~\cite{seki-07}, 
the apparent exponent is expected to crossover from
$z=2$ to $z=4$.}.

\begin{figure}[t]
\begin{center}
\resizebox{0.9\columnwidth}{!}{
\includegraphics{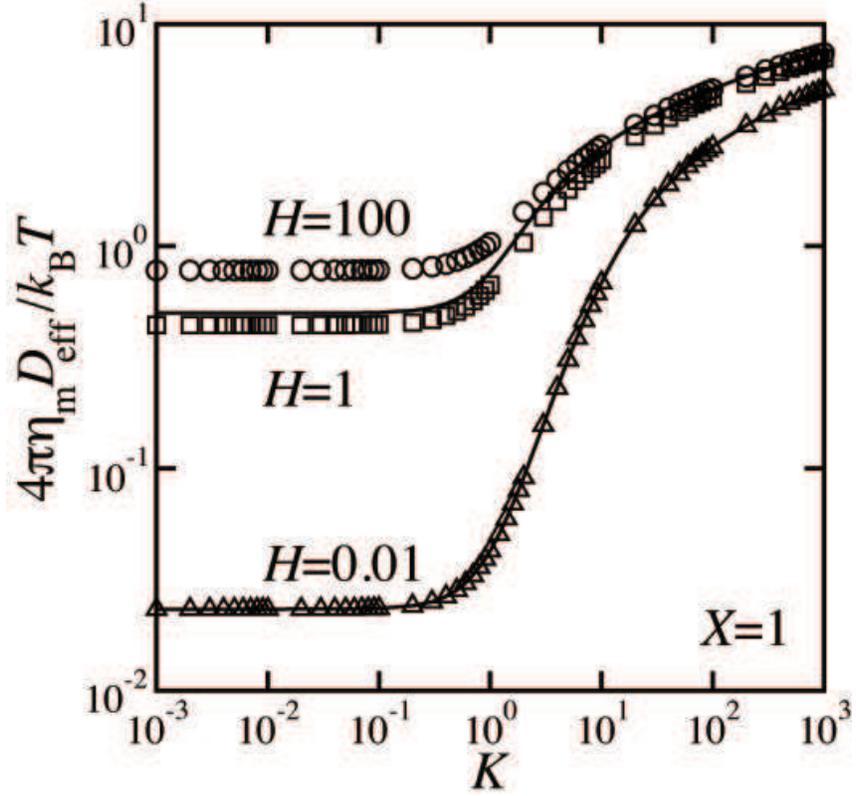}
}
\caption{
Scaled effective diffusion coefficient $4\pi\eta_{\rm m}D_{\rm eff}/k_{\rm B}T$ 
as a function of dimensionless wavenumber $K=k/\nu$.
The dimensionless correlation length, $X=\xi \nu$, is fixed to unity,
while $H=h \nu$ is the dimensionless
distance between
the membrane and the wall and its value varies between 0.01 to 100.
Symbols are numerical calculations and the solid lines correspond to the
analytical expression given in Ref.~\cite{seki-07} obtained in the limit of
small $H$. 
}
\label{fig:effdiff}
\end{center}
\end{figure}

In another study, Ramachandran \textit{et al.}~\cite{sanoop-conc-10} 
used the general mobility tensor of
Eq.~(\ref{eqn:cothoseen}) to numerically calculate the effective diffusion
coefficient.
This quantity includes several parameters such as the correlation length
$\xi$, the hydrodynamic screening length $\nu^{-1}$, and the distance $h$
between the membrane and the wall.
Their main findings are reproduced in Fig.~\ref{fig:effdiff} where we plot the
computed effective diffusion
coefficient $D_{\rm eff}$ as a function of the dimensionless wavenumber
$K=k/\nu$.
The dimensionless quantity $X=\xi\nu=1$ is held fixed,
while $H=h\nu$ is  varied between three representative
values: $H=0.01, 1, 100$.
The symbols correspond to the numerical estimates and the solid lines represent the
analytical expression derived in Ref.~\cite{seki-07}.
For $K \ll 1$, we see that $D_{\rm eff}$ is almost a constant that
depends on $H$, while it increases logarithmically for $K \gg 1$.
Notice that this logarithmic dependence is a characteristic feature specific to
2d fluid membranes, and does not exist for 3d critical fluids~\cite{kawasaki-70}.

%%%%%%%%%%%%%%%%%%%%%%%%%%%%%%%%%%%%%%%%%%%%%%%%%%%%%%%%%%%%%%%%%%%%%%%%%
\section{2d microemulsion model}
\label{sec:microemulsion}

\subsection{Hybrid lipids}
%%%%%%%%%%%%%%%%%%%%%%%%%%%%%%%%%%%%%%%%%%%%%%%%%%%%%%%%%%%%%%%%%%%%%%%%%

One central issue in regards to lipid rafts in biomembranes is to identify
the physical mechanism that determines the finite domain size.
In this section, we discuss the possibility of having another length scale
that is associated with rafts, and which is
different from the correlation length $\xi$ discussed so far.

In Sec.~\ref{sec:phase} we mentioned various ternary lipid mixtures,
such as DOPC/DPPC/ cholesterol that have been extensively studied.
If we take a closer look at the DOPC and DPPC molecules, we note that
for DOPC the two chains contain each
one unsaturated bond (doubly unsaturated lipid),
while for DPPC both  chains are fully saturated.
Other type of lipids such as POPC\footnote{POPC: palmitoyl-oleoyl-phosphatidylcholine}
and SOPC\footnote{SOPC: stearoyl-oleoyl-phosphatidylcholine} has one
unsaturated chain and one saturated chain. They are sometimes called {\it "hybrid lipids"}.
Real biomembranes contain a much larger fraction of hybrid lipids than doubly unsaturated
ones such as DOPC~\cite{vanmeer-08}.

Considering a mixture of doubly unsaturated lipid, saturated
lipid, and hybrid lipid, we realize that this system resembles {\it microemulsions} ---
a well-studied 3d ternary
liquid mixture composed of water/oil/surfactant.
In microemulsions, surfactants are absorbed at water/oil interfaces and reduce
the water/oil interfacial tension~\cite{gompper-schick}.
By analogy, we expect that hybrid lipids will play a similar role in 2d by absorbing at interfaces between
the doubly-unsaturated lipids and the saturated lipids, in order to reduce the line
tension.
This effect can be referred to as  {\it "line activity"} in 2d systems, and is similar to
{\it "surface activity"} of surfactants in 3d.

A lattice model for 
this type of 2d microemulsion-like ternary mixtures has been proposed
by Brewster \textit{et al.}~\cite{brewster-09,brewster-10}. In their model they 
calculated the reduction of the line tension due to the presence
of hybrid lipid in small quantities.
This lattice model for ternary mixtures has been recently further extended
by Palmieri\textit{et al.}~\cite{palmieri-13a,palmieri-13b} for any fraction of unsaturated, saturated and
hybrid lipids.
It was shown that the correlation length of concentration fluctuations decreases
dramatically with increasing amounts of hybrid lipids, and 
nanoscale fluctuations are more probable in the presence of hybrid lipids.
In their second work~\cite{palmieri-13b}, Palmieri \textit{et al.} concluded that hybrid lipids
increase the lifetime of fluctuations.

Yamamoto et al. \textit{et al.}~\cite{yamamoto-10,yamamoto-11} showed that,
in some cases,  the reduction of  line
tension is more pronounced than for three-component lipid mixtures.
More details are provided by Palmieri \textit{et al.} in a separate
contribution to this special issue.

%%%%%%%%%%%%%%%%%%%%%%%%%%%%%%%%%%%%%%%%%%%%%%%%%%%%%%%%%%%%%%%%%%%%%%%%%
\subsection{Coupling between concentration field and orientation field}
%%%%%%%%%%%%%%%%%%%%%%%%%%%%%%%%%%%%%%%%%%%%%%%%%%%%%%%%%%%%%%%%%%%%%%%%%

\begin{figure}[t]
\begin{center}
\resizebox{0.9\columnwidth}{!}{
\includegraphics{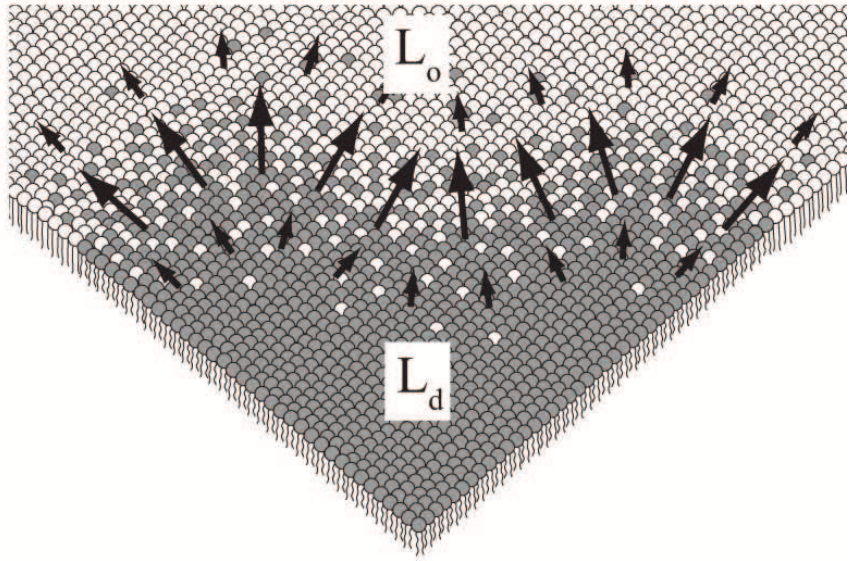}
}
\caption{
Lipid monolayer consisting of saturated lipids and hybrid lipids.
The orientational vector field $\mathbf{m}$ points from the
L$_{\rm d}$-phase to the L$_{\rm o}$-phase, and its magnitude becomes
large at the interface between the two phases.
}
\label{fig:vector}
\end{center}
\end{figure}

Hirose \textit{et al.}~\cite{hirose-09,hirose-12} extended the models by 
Brewster \textit{et al.}~\cite{brewster-09,brewster-10} and 
Yamamoto \textit{et al.}~\cite{yamamoto-10,yamamoto-11} first to lipid 
monolayers and then to coupled bilayers as will be discussed later. 
In their work they proposed a Ginzburg--Landau model for systems 
containing hybrid lipids in addition to saturated ones.
They introduced a 2d orientational vector-field $\mathbf{m}({\mathbf r})$,
which points from the unsaturated chain to the saturated chain of a
hybrid lipid.
As shown in Fig.~\ref{fig:vector}, the orientational vector  $\mathbf{m}$ is aligned
toward the L$_{\rm o}$-phase, and its magnitude increases at the interface.
In addition to Eq.~(\ref{eqn:freeE}), the  free
energy has some additional terms~\cite{hirose-12}:
\begin{equation}
F [\mathbf{m},\phi] = \int {\rm d}{\mathbf r} \,
\left[ \frac{E}{2}(\nabla \cdot \mathbf{m})^2 +
\frac{b}{2} \mathbf{m}^2- \Lambda \mathbf{m} \cdot (\nabla \phi)
\right],
\label{freeenergyhybrid}
\end{equation}
where $E$ is the 2d elastic constant, and both $b$ and $\Lambda$
are positive coefficients.

The total free energy is the sum of Eqs.~(\ref{eqn:freeE}) and (\ref{freeenergyhybrid})
can be minimized with respect to $\mathbf{m}({\mathbf r})$. It then results in 
effective monolayer free-energy that depends only $\phi$, and  within the long-wavelength
approximation the expression is:
\begin{equation}
F_{\rm m}[\phi] = \int {\rm d}{\mathbf r} \,
\left[ \frac{B}{2} ( \nabla^2 \phi )^2 - \frac{A}{2} ( \nabla \phi )^2
+ \frac{a}{2} \phi^2 + \frac{1}{4}\phi^4 \right],
\label{mono_minimized}
\end{equation}
where $B=E \Lambda^2/b^2$ and $A=\Lambda^2/b-c$.
When the coupling coefficient $\Lambda$ is large enough, $A>0$, the
above effective free-energy exhibits an instability at a finite
wavenumber $k^{\ast} = \sqrt{A/2B}$~\cite{brazovskii}.

The 2d free-energy in Eq.~(\ref{mono_minimized}) has the same form as that
for 3d microemulsions~\cite{gompper-schick}.
The important dimensionless parameter is
$\theta= -A / \sqrt{4 a B}$ valid for $a>0$ ($T > T_{\rm c}$).
When $\vert \theta \vert < 1$, the correlation function,
$G_{\phi \phi}(\mathbf{r})=\langle \delta \phi(\mathbf{r}) \delta \phi(0)\rangle$,
obtained from Eq.~(\ref{mono_minimized}) decays with an oscillatory
component.
The correlation length and the period length are given by
$\xi = (4B/a)^{1/4} (1 + \theta)^{-1/2}$ and
$\lambda/2 \pi = (4B/a)^{1/4} (1 - \theta)^{-1/2}$, respectively.
Although both lengths are finite for $\vert \theta \vert <1$, $\xi$
diverges at $\theta=-1$ and $\lambda$ at $\theta=1$.
When the correlation length $\xi$ diverges, micro-phase separated structures
such as the stripe  or  hexagonal phases become more
stable~\cite{gompper-schick}.

%%%%%%%%%%%%%%%%%%%%%%%%%%%%%%%%%%%%%%%%%%%%%%%%%%%%%%%%%%%%%%%%%%%%%%%%%
\subsection{Curvature instability}
%%%%%%%%%%%%%%%%%%%%%%%%%%%%%%%%%%%%%%%%%%%%%%%%%%%%%%%%%%%%%%%%%%%%%%%%%

Another physical mechanism which leads to very similar microemulsion-like
free-energy is the curvature instability~\cite{LA}.
In this model, the concentration field is coupled to the membrane curvature.
Within the Monge representation, the shape of a membrane can be described
by its height $h({\mathbf r})$, and the curvature elasticity energy,
Eq.~(\ref{curvature}), is given approximately by:
\begin{equation}
F [h,\phi] = \int {\rm d}{\mathbf r} \,
\left[ \frac{\kappa}{2}(\nabla^2 h)^2 + \frac{\Sigma}{2}(\nabla h)^2
- \Upsilon (\nabla^2 h) \phi
\right],
\label{monge}
\end{equation}
where $\kappa$ is the bending rigidity introduced earlier, $\Sigma$ the
membrane surface tension, and $\Upsilon$ is a coupling
coefficient between local composition and curvature, while the term 
associated with the Gaussian curvature is neglected.

Physically speaking, the coupling term represents a spontaneous curvature
that depends on the local concentration $\phi$.
By adding Eqs.~(\ref{eqn:freeE}) and (\ref{monge}), one can minimize the
total free-energy with respect to $h({\mathbf r})$, and obtain a
free-energy similar to Eq.~(\ref{mono_minimized}).
Equilibrium shapes of modulated vesicles were investigated in detail in two limits:
strong segregation limit (temperatures much smaller than the critical 
temperature)~\cite{AKK,KAKT}, and  weak segregation one (close to the 
critical temperature)~\cite{TKAK}.

For bilayers, in particular, the membrane curvature can be naturally 
coupled to the compositional asymmetry between the two leaflets, and it 
can also lead to a curvature instability~\cite{SPAM,MS,KK93,KGL}.
From the energetic point of view, the frustration to form bilayers out of two
monolayers can be avoided by creating a composition asymmetry in binary
membranes.
Using this idea, Schick and co-workers proposed a free energy similar to
Eq.~(\ref{mono_minimized}) for bilayer membranes~\cite{schick-12,RS13}, while
Meinhardt \textit{et al.}~\cite{MVS} considered the coupling effect between the
curvature and the membrane thickness, which also results in the formation
of modulated phases at low temperatures.

%%%%%%%%%%%%%%%%%%%%%%%%%%%%%%%%%%%%%%%%%%%%%%%%%%%%%%%%%%%%%%%%%%%%%%%%%
\subsection{Coupled modulated monolayers}
%%%%%%%%%%%%%%%%%%%%%%%%%%%%%%%%%%%%%%%%%%%%%%%%%%%%%%%%%%%%%%%%%%%%%%%%%

\begin{figure}[t]
\begin{center}
\resizebox{0.9\columnwidth}{!}{
\includegraphics{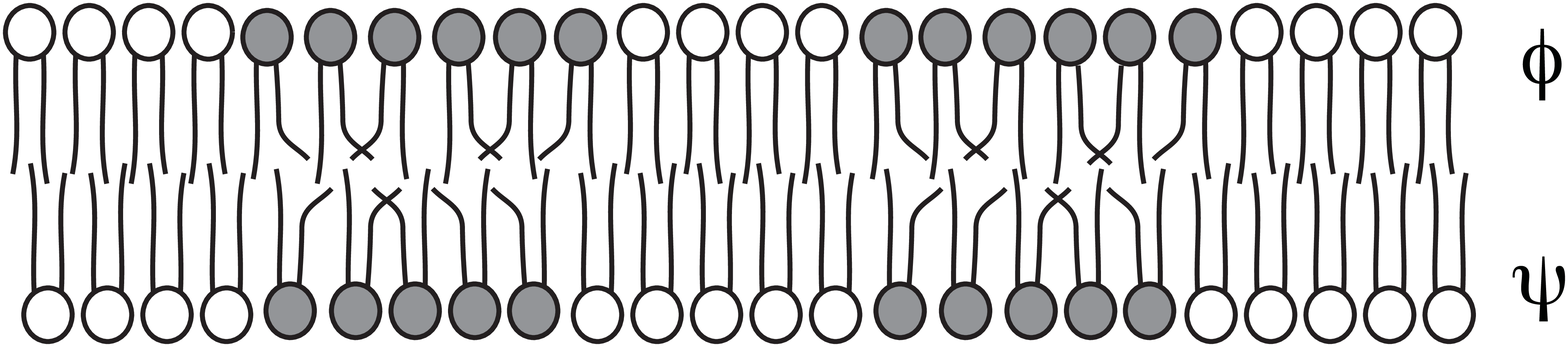}
}
\caption{
Schematic illustration of two coupled modulated monolayers of concentration
$\phi$ (upper) and $\psi$ (lower) forming a bilayer membrane.
Each monolayer is composed of a binary mixture of saturated lipid (white) and 
hybrid lipid (gray), which
can have a lateral modulation in $\phi$ and $\psi$.
}
\label{fig:cmb}
\end{center}
\end{figure}

Hirose \textit{et al.}~\cite{hirose-09,hirose-12} considered bilayers
composed of two modulated monolayers whose free energies are described by
Eq.~(\ref{mono_minimized}).
Denoting the concentrations of the upper and lower
leaflets by $\phi$ and $\psi$, respectively (see Fig.~\ref{fig:cmb}),
the total bilayer free-energy is given by

\begin{align}
F_{\rm b}[\phi, \psi ] = \int {\rm d}{\mathbf r} \,
\biggl[ & \frac{B_\phi}{2}( \nabla^2 \phi)^2 - \frac{A_\phi}{2}(\nabla \phi)^2
+ \frac{a_{\phi}}{2}\phi^2 + \frac{1}{4}\phi^4
\nonumber \\
+ & \frac{B_\psi}{2}( \nabla^2 \psi)^2 - \frac{A_\psi}{2}( \nabla \psi)^2
+ \frac{a_{\psi}}{2}\psi^2 + \frac{1}{4}\psi^4
- \Xi \phi \psi \biggr],
\label{freeenergy}
\end{align}
and is constructed from Eq.~(\ref{mono_minimized}) for each monolayer.
The coupling between the two leaflets is taken into account by the
last term in which $\Xi$ is the coupling
coefficient~\cite{collins-08,May,CK,WLM,PS}.

Above the transition temperature, both static and dynamic properties
of concentration fluctuations have been investigated in Refs.~\cite{hirose-09,hirose-12}.
For example, the static partial structure factor
$S_{\phi \phi}(\mathbf{k})=\langle \delta \phi(\mathbf{k})
\delta \phi (-\mathbf{k}) \rangle$ can be obtained by the 2d Fourier
transform of the correlation function $G_{\phi \phi}(\mathbf{r}) =
\langle \delta \phi (\mathbf{r}) \delta \phi(0) \rangle$ as introduced
before.
Similarly, the other partial structure factors,
$S_{\psi \psi}(\mathbf{k})$ and $S_{\phi \psi}(\mathbf{k})$, can be obtained.

\begin{figure}[t]
\begin{center}
\resizebox{0.7\columnwidth}{!}{
\includegraphics{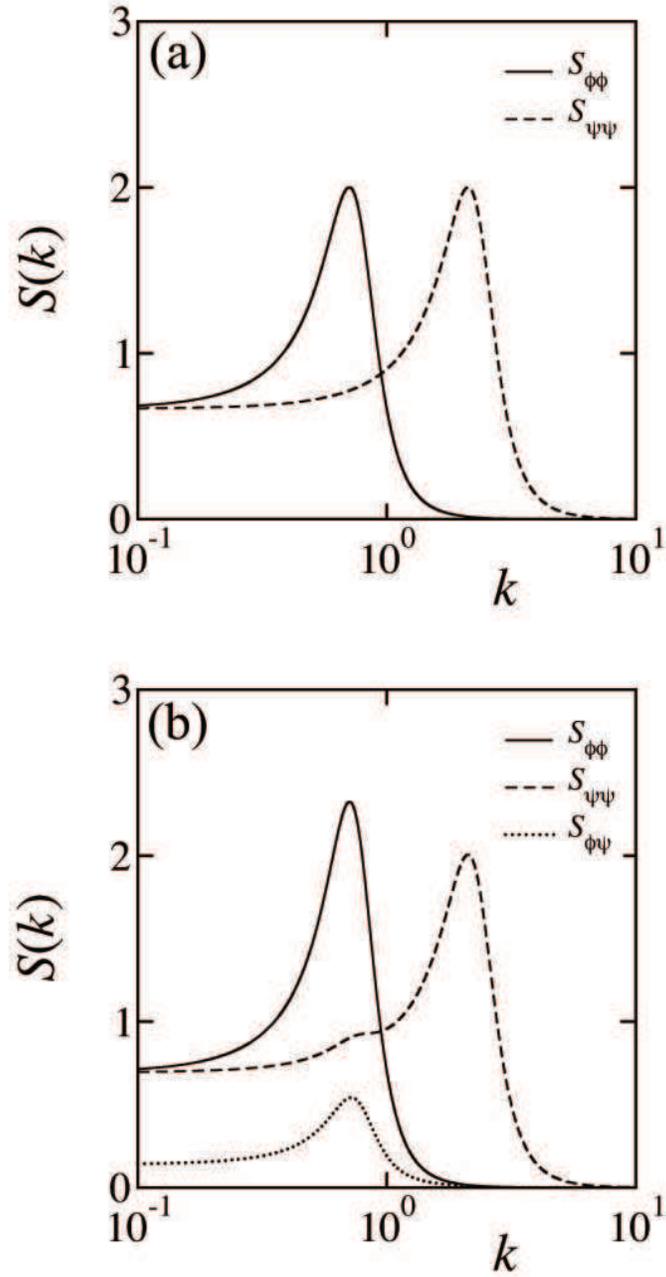}
}
\caption{Bilayer structure factors $S_{\phi\phi}$, $S_{\psi\psi}$, and
$S_{\phi\psi}$ as a function of the wavenumber $k$.
(a) The decoupled case, $\Xi= 0$. (b) The coupled case with
$\Xi =0.3$.
}
\label{fig:cmbsk}
\end{center}
\end{figure}

In Fig.~\ref{fig:cmbsk}(a) and (b), we plot the structure factors of
the decoupled and coupled cases, respectively.
As an illustration of the coupling effect, we consider that
the $\phi$ and $\psi$ leaflets have different
characteristic wavenumbers~\cite{palmieri-13a}: $k_{\phi}^* <  k_{\psi}^*$, while
the heights of the two peaks are set to be equal.
The peak height of $S_{\phi\phi}$ at $k_{\phi}^*$ is increased due to
the coupling effect, whereas that of $S_{\psi\psi}$ at $k_{\psi}^*$
is almost unchanged, as compared with the decoupled case.
We also plot $S_{\phi\psi}$ that represents the cross-correlation of
fluctuations between the two leaflets.
This quantity is proportional to the coupling constant $\Xi$ and its peak
position is essentially determined by that of $S_{\phi\phi}$ at
$k_{\phi}^*$.

The dynamical fluctuations in composition, $\delta \phi (\mathbf{r},t)$
and $\delta \psi (\mathbf{r},t)$, are also considered for coupled
modulated monolayers~\cite{hirose-09,hirose-12}.
Since the exchange of lipids between the two monolayers is negligible,
the time evolution of $\delta \phi$ and $\delta \psi$ are given by the
diffusive equations (in the absence of any hydrodynamic effects).
In the decoupled case, one can show that the $S_{\phi \phi}$ and 
$S_{\psi \psi}$ structure factors decay with a single exponential 
characterized by a decay rate that depends on the wavenumber.
For nonzero coupling coefficient, $\Xi \neq 0$, it was shown that the
decay of concentration fluctuations is described by a sum of two
exponentials.
Generally speaking, the coupling affects the decay time of the leaflet with
the smaller wavenumber (larger length scale).

Below the transition temperature when both monolayers exhibit micro-phase
separation and form either stripe (S) or hexagonal (H) phase, the leaflet
coupling brings about various combinations of the monolayer modulated
phases.
When the two leaflets have the same preferred periodicity,
$k_{\phi}^* = k_{\psi}^*$, Hirose \textit{et al.}~\cite{hirose-09} obtained the mean-field
phase diagram exhibiting various combinations of modulated
structures.
In some cases, the periodic structure in one of the monolayers induces
a similar modulation in the second monolayer.
Moreover, the region of the induced modulated phase expands as the coupling
parameter $\Xi$ becomes larger.

\begin{figure}[t]
\begin{center}
\resizebox{0.9\columnwidth}{!}{
\includegraphics{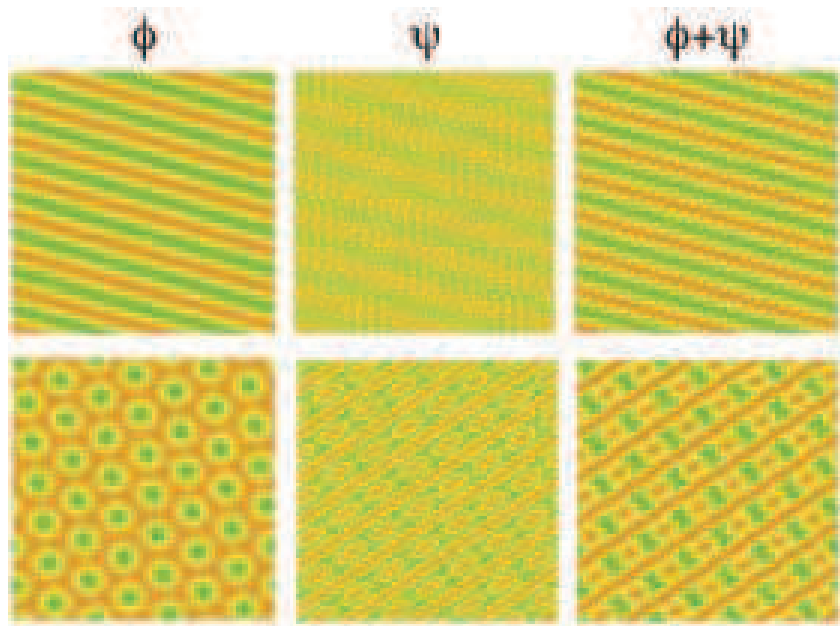}
}
\caption{
Patterns of two coupled modulated leaflets of concentration $\phi$ and $\psi$.
In the three top parts the $\phi$ and $\psi$ leaflets consist of a stripe phase, 
while  in the three bottom parts the coupling is between
hexagonal (in $\phi$) and stripe (in $\psi$) phases, each having 
a different characteristic wavelength.
Left: $\phi$-monolayer, middle: $\psi$-monolayer, and right: $\phi+\psi$. 
}
\label{fig:cmbpattern}
\end{center}
\end{figure}

When the preferred periodicity of the two leaflets is different,
$k_{\phi}^* \neq k_{\psi}^*$, it is necessary to 
solve numerically the coupled time-evolution equations given by
\begin{align}
& \frac{\partial \phi}{\partial t} =
L_{\phi} \nabla^2 \frac{\delta F_{\rm b}}{\delta \phi},
\nonumber \\
& \frac{\partial \psi}{\partial t} =
L_{\psi} \nabla^2 \frac{\delta F_{\rm b}}{\delta \psi},
\label{conc_evol_s}
\end{align}
where the bilayer free-energy, $F_{\rm b}[\phi,\psi]$, is given by
Eq.~(\ref{freeenergy}).
It was shown~\cite{hirose-09,hirose-12} that various complex patterns are formed due to the frustration
between the two incommensurate modulated structures.
In Fig.~\ref{fig:cmbpattern}, we show examples of complex patterns created
from two stripe structures (top), or stripe and hexagonal structures (bottom).
Broadly speaking, the structure with the larger wavelength dominates when
the coupling is large enough, which is in accord with the properties of the
static structure factors.
More details are discussed in Refs.~\cite{hirose-09,hirose-12}.

%%%%%%%%%%%%%%%%%%%%%%%%%%%%%%%%%%%%%%%%%%%%%%%%%%%%%%%%%%%%%%%%%%%%%%%%%
\section{Outlook}
%%%%%%%%%%%%%%%%%%%%%%%%%%%%%%%%%%%%%%%%%%%%%%%%%%%%%%%%%%%%%%%%%%%%%%%%%

In this article, we reviewed some of the more recent
physical and chemical-physics studies
concerning the static and dynamic properties of lateral phase-separation
in multi-component lipid bilayers.
We intentionally avoided including other studies based on biological
perspectives as we preferred to keep this review focused on physical
concepts and their impact on the understanding of biomembranes.

We have shown that even for the simplified case of lipid membranes composed
of only three components, the inhomogeneity in the lateral composition
couples with the membrane shape, and can lead to a rich variety of interesting
phenomena.
We believe that purely physical phenomena such as phase separation and
diffusion that occur in biomembranes may play an important role in biological
systems.
Since these physical phenomena can be described using appropriate and well-defined 
physical models, more quantitative arguments are possible regarding the static and
dynamic features of lipid domains and, presumingly, with an impact on rafts.

It will be of interest to explore in the future the
phase separation in multi-component lipid membranes 
in presence of glycolipids
and membrane proteins, because the latter are abundant in biomembranes~\cite{kusumi}.
Understanding the interactions between protein molecules embedded inside multi-component membranes~\cite{RD08},
as well as between different multi-component membranes would be also
of interest for future investigations~\cite{AK03}.

%%%%%%%%%%%%%%%%%%%%%%%%%%%%%%%%%%%%%%%%%%%%%%%%%%%%%%%%%%%%%%%%%%%%%%%%%
\section*{Acknowledgments}
%%%%%%%%%%%%%%%%%%%%%%%%%%%%%%%%%%%%%%%%%%%%%%%%%%%%%%%%%%%%%%%%%%%%%%%%%

Some of the works reviewed in this article have been conducted together
with our collaborators; G. Gompper, Y. Hirose, M. Imai,
Y. Kanemori, S. Ramachandran, Y. Sakuma, K. Seki, N. Shimokawa and
M. Yanagisawa.
We greatly acknowledge their contributions.
We also thank H. Diamant, Y. Fujitani,  T. Hamada, T. Kato, P. B. Sunil Kumar,
S. L. Keller, C.-Y. D. Lu, N. Oppenheimer, B. Palmieri, E. Sackmann,
S. A. Safran, M. Schick and T. Yamamoto for useful discussions.

SK would like to acknowledge support from the Grant-in-Aid for Scientific
Research on Innovative Areas {\it "Fluctuation \& Structure"} (No. 25103010),
grant No. 24540439 from the MEXT (Japan), and the JSPS Core-to-Core
Program {\it "International research network for nonequilibrium dynamics
of soft matter"}.
DA acknowledges support from the US-Israel Binational Science Foundation
(BSF) under Grant No. 2012/060 and from the Israel Science Foundation (ISF)
under Grant No. 438/12.

%%%%%%%%%%%%%%%%%%%%%%%%%%%%%%%%%%%%%%%%%%%%%%%%%%%%%%%%%%%%%%%%%%%%

%%%%%%%%%%%%%%%%%%%%%%%%%%%%%%%%%%%%%%%%%%%%%%%%%%%%%%%%%%%%%%%%%%%%
%fig 1
%%%%%%%%%%%%%%%%%%%%%%%%%%%%%%%%%%%%%%%%%%%%%%%%%%%%%%%%%%%%%%%%%%%%%%%%%

%%%%%%%%%%%%%%%%%%%%%%%%%%%%%%%%%%%%%%%%%%%%%%%%%%%%%%%%%%%%%%%%%%%%%%%%%%
%fig 2
%%%%%%%%%%%%%%%%%%%%%%%%%%%%%%%%%%%%%%%%%%%%%%%%%%%%%%%%%%%%%%%%%%%%%%%%%

%%%%%%%%%%%%%%%%%%%%%%%%%%%%%%%%%%%%%%%%%%%%%%%%%%%%%%%%%%%%%%%%%%%%%%%%%
%fig 3
%%%%%%%%%%%%%%%%%%%%%%%%%%%%%%%%%%%%%%%%%%%%%%%%%%%%%%%%%%%%%%%%%%%%%%%%%

%%%%%%%%%%%%%%%%%%%%%%%%%%%%%%%%%%%%%%%%%%%%%%%%%%%%%%%%%%%%%%%%%%%%%%%%%
%fig 4
%%%%%%%%%%%%%%%%%%%%%%%%%%%%%%%%%%%%%%%%%%%%%%%%%%%%%%%%%%%%%%%%%%%%%%%%%

%%%%%%%%%%%%%%%%%%%%%%%%%%%%%%%%%%%%%%%%%%%%%%%%%%%%%%%%%%%%%%%%%%%%%%%%%
%fig 5
%%%%%%%%%%%%%%%%%%%%%%%%%%%%%%%%%%%%%%%%%%%%%%%%%%%%%%%%%%%%%%%%%%%%%%%%%

%%%%%%%%%%%%%%%%%%%%%%%%%%%%%%%%%%%%%%%%%%%%%%%%%%%%%%%%%%%%%%%%%%%%%%%%%
%fig 6
%%%%%%%%%%%%%%%%%%%%%%%%%%%%%%%%%%%%%%%%%%%%%%%%%%%%%%%%%%%%%%%%%%%%%%%%%

%%%%%%%%%%%%%%%%%%%%%%%%%%%%%%%%%%%%%%%%%%%%%%%%%%%%%%%%%%%%%%%%%%%%%%%%%
%fig 7
%%%%%%%%%%%%%%%%%%%%%%%%%%%%%%%%%%%%%%%%%%%%%%%%%%%%%%%%%%%%%%%%%%%%%%%%%

%%%%%%%%%%%%%%%%%%%%%%%%%%%%%%%%%%%%%%%%%%%%%%%%%%%%%%%%%%%%%%%%%%%%%%%%%
%fig 8
%%%%%%%%%%%%%%%%%%%%%%%%%%%%%%%%%%%%%%%%%%%%%%%%%%%%%%%%%%%%%%%%%%%%%%%%%

%%%%%%%%%%%%%%%%%%%%%%%%%%%%%%%%%%%%%%%%%%%%%%%%%%%%%%%%%%%%%%%%%%%%%%%%%
%fig 9
%%%%%%%%%%%%%%%%%%%%%%%%%%%%%%%%%%%%%%%%%%%%%%%%%%%%%%%%%%%%%%%%%%%%%%%%%

%%%%%%%%%%%%%%%%%%%%%%%%%%%%%%%%%%%%%%%%%%%%%%%%%%%%%%%%%%%%%%%%%%%%%%%%%
%fig 10
%%%%%%%%%%%%%%%%%%%%%%%%%%%%%%%%%%%%%%%%%%%%%%%%%%%%%%%%%%%%%%%%%%%%%%%%%

%%%%%%%%%%%%%%%%%%%%%%%%%%%%%%%%%%%%%%%%%%%%%%%%%%%%%%%%%%%%%%%%%%%%%%%%%
%fig 11
%%%%%%%%%%%%%%%%%%%%%%%%%%%%%%%%%%%%%%%%%%%%%%%%%%%%%%%%%%%%%%%%%%%%%%%%%

%%%%%%%%%%%%%%%%%%%%%%%%%%%%%%%%%%%%%%%%%%%%%%%%%%%%%%%%%%%%%%%%%%%%%%%%%
%fig 12
%%%%%%%%%%%%%%%%%%%%%%%%%%%%%%%%%%%%%%%%%%%%%%%%%%%%%%%%%%%%%%%%%%%%%%%%%

%%%%%%%%%%%%%%%%%%%%%%%%%%%%%%%%%%%%%%%%%%%%%%%%%%%%%%%%%%%%%%%%%%%%%%%%%
%fig 13
%%%%%%%%%%%%%%%%%%%%%%%%%%%%%%%%%%%%%%%%%%%%%%%%%%%%%%%%%%%%%%%%%%%%%%%%%

%%%%%%%%%%%%%%%%%%%%%%%%%%%%%%%%%%%%%%%%%%%%%%%%%%%%%%%%%%%%%%%%%%%%%%%%%
%fig 14
%%%%%%%%%%%%%%%%%%%%%%%%%%%%%%%%%%%%%%%%%%%%%%%%%%%%%%%%%%%%%%%%%%%%%%%%%

%%%%%%%%%%%%%%%%%%%%%%%%%%%%%%%%%%%%%%%%%%%%%%%%%%%%%%%%%%%%%%%%%%%%%%%%%
%fig 15
%%%%%%%%%%%%%%%%%%%%%%%%%%%%%%%%%%%%%%%%%%%%%%%%%%%%%%%%%%%%%%%%%%%%%%%%%

\end{document}